\documentclass{article}

\usepackage{graphicx,amsmath,amssymb,wrapfig,color,amsfonts}
\usepackage[section]{placeins}
\usepackage{url,enumitem}
\usepackage[]{hyperref}

\newcommand{\onehalf}{\frac{1}{2}}
\newcommand{\pr}[1]{{#1}_m}
\newcommand{\rhot}{\widetilde{\rho}}
\newcommand{\rhoh}{\widehat{\rho}}
\newcommand{\ut}{\widetilde{u}}
\newcommand{\uh}{\widehat{u}}
\newcommand{\wt}{\widetilde{w}}
\newcommand{\wh}{\widehat{w}}

\newcommand{\commentout}[1]{}

\date{}

\begin{document}

\title{Air-burst Generated Tsunamis}
\author{Marsha Berger\footnote{Courant Institute, New York University, 251 Mercer St.,
NY, NY 10012}  \hspace{1in} Jonathan Goodman$^*$}


\maketitle

\begin{abstract}
This paper examines the questions of whether smaller 
asteroids that burst in the air over water 
can generate tsunamis that could pose a threat to distant 
locations.
Such air burst-generated tsunamis are qualitatively different than the
more frequently studied earthquake-generated tsunamis,
and differ as well from impact asteroids.
Numerical simulations are presented using the shallow water equations in
several settings, demonstrating very little tsunami threat from this scenario. 
A model problem with an explicit  solution
that demonstrates and explains the same phenomena 
found in the computations is analyzed.
We discuss the question of whether compressibility and dispersion are
important effects that should be included, and show results from
a more sophisticated model problem using the linearized Euler equations 
that begins to addresses this. 
\end{abstract}

\hspace*{.16in}{\bf Keywords:}{ tsunami;  asteroid-generated air-burst;  shallow water
equations;\\
\hspace*{.38in}linearized Euler equations.}


\section{Introduction}\label{sec:intro}

In Feb. 2013, an asteroid with a 20 meter diameter 
burst 30 km high in the atmosphere over Chelyabinsk, causing substantial 
local damage over a 20,000 $\text{km}^2$ region \cite{Popova2013}.
The question arises, what would be  the effect of an  asteroid that 
bursts over the ocean instead of land? 
The concern is that the atmospheric
blast wave might generate a tsunami threatening  populated coastlines far away.

There is little literature on air-burst-generated tsunamis. Most
of the literature on asteroids study the more complicated
case of water impacts, 
where the meteorite splashes into the ocean 
\cite{WeissWunnemannBahlburg:2006,GislerWeaverGitting:2010,Gisler:2008}.
This involves much more complicated physics. The only reference we are aware of
that relates to a blast-driven water wave is from the  1883 volcanic explosion of
Krakatoa  \cite{krakatoa}. The authors report a tide
gauge in San Francisco registered a wave  that could not be explained by
a tsunami.  
There is also some analytic work in \cite{krantzer:keller},
where they derive asymptotic formulas for water waves from explosions and
from initial cavities.
There is more
literature on meteo-tsunamis. These are also driven by
air-pressure events and have similarities to our case, but occur in a
different regime of air speed and water depth.

This paper studies the behavior of air-burst generated tsunamis, 
to better understand the potential threat. 
In the first part of the paper, we present simulations 
under a range of conditions using the shallow water equations and the
GeoClaw software package \cite{geoclaw:URL}. 
We compute  the ocean's response to an  overpressure
as calculated in \cite{aftosmis:bolide2016}.
The overpressure was found by  simulating
the blast wave in air, and  extracting the ground footprint.
Roughly speaking, the blast wave model corresponds to the largest meteor that
deposits all of its energy in the atmosphere without actually reaching
the water surface.
With this forcing, if there
is no sizeable response then we can conclude that that air-bursts do
not effectively transfer energy to the ocean, and there is little
threat of distant inundation. 

Typically the shallow water equations are
used for long-distance propagation, since they efficiently
and affordably propagate waves over large trans-oceanic distances.
Other alternatives, such as the Boussinesq equations,
are much more expensive, and at least for the 
case of earthquake-generated tsunamis the difference seems to be small
\cite{Liu:chap9}.

In general, our
results using the shallow water equations suggest that
air-burst generated tsunamis are too small to cause
much coastal damage.
Of course, depending on local bathymetry
there could an unusual response that is significant. For example, 
Crescent City, California is well-known to be subject to inundation due to the
configuration of its harbor and local bathymetry. 
However, we find that to generate a
large enough response  so that  the water floods the coastline,
the blast has to be so close that the blast itself is the more
dangerous phenomenon.  This is also the conclusion reached by
Gisler et al. \cite{GislerWeaverGitting:2010} and Melosh
\cite{Melosh:2003}  for the case of 
asteroid water impacts.

In the second part of this paper, we study model problems 
to better understand and describe the phenomena we compute
in the first part. 
The first model problem is based on 
the one-dimensional shallow water equations
for which we can obtain an explicit closed form solution.
It assumes a traveling wave form for the pressure forcing.
Actual blast waves only approximately satisfy this 
hypothesis for a short time before their amplitudes decay. 
Nevertheless, the model
explains several key features that we observe in the 
two-dimensional simulations. 
We observe  a response wave that moves with the speed of the atmospheric 
forcing. There is also the gravity wave, or tsunami, moving at the
shallow water wave speed, that is  
generated by the initial transient of
the atmospheric forcing.  We study in 
detail the response  wave, or `forced' wave,  but the two are closely related.
The analysis shows that the forced wave is proportional to the local depth
of the water at each location, a phenomena clearly seen in our 
computations. 
The model problem also allows us to assess the importance of 
nonlinear modeling. For 
most physical situations related to air-burst tsunamis,
the linear and nonlinear models give similar predictions.

In our final section
we assess the effect of corrections
to the shallow water equations arising from compressibility and
dispersion using
a second model problem - the linearized Euler 
equations.   
Air bursts have a much shorter time scale than 
earthquake-generated tsunamis, comparable to the acoustic
travel time to the ocean floor. This leads to the
question of whether compressibility of the ocean water 
could be a significant factor. 
In addition, air bursts have  much shorter wavelengths, 
on the order of 10 to 20 kilometers, at least for meteors with
diameter less than 200 meters or so.
Recall that the shallow water model 
results from  assuming long wavelengths 
and incompressibility of the water. 
Our results show that for air-burst generated tsunamis,
dispersion can be significant but that compressibility is less so, suggesting
interesting avenues for future work.

This work is an outgrowth of the 2016 NASA-NOAA Asteroid-generated Tsunami
and Associated Risk Assessment Workshop. 
The workshop conclusions are summarized in \cite{AGT:NASATM}. Several other
researchers also performed simulations, and videos of
all talks are available on-line\footnote{All
presentations are available at
https://tsunami-workshop.arc.nasa.gov/workshop2016/sched.php}.

\section{Two-dimensional Simulations}\label{sec:sims}

In this section we present results from two sets of simulations. 
We use a 250MT blast, which roughly
corresponds to a meteor with a 200 meter diameter 
entering the atmosphere with a speed of 20 km/sec.
Generally speaking, this is the
largest asteroid that would not splash into the 
water.\footnote{
Initially we used a blast wave corresponding to a 100MT blast, but since
no significant response was found we do not include those results here.
We also did simulations where we
increased the pressure forcing by a factor 
of 1.2 with no change to the conclusions.}  
For each location, we did several simulations varying the blast
locations with no meaningful difference in results, so we only present one
representative computation in each set of simulations.

\commentout{
We locate the 100MT blast in the middle of
the South China Sea. This spot was chosen since the
coastlines surrounding it have a great variety of bathymetry,
including extensive  shallow continental shelfs (near Malaysia and
China) and places with almost
no shelf (near the Philipines and Vietnam).  It was
also studied in the 2015 Planetary Defense conference as part of the 
Table Top exercise. We show results near the center of the blast and at
several points near shore. We find that by the time the waves reach shore, their
amplitude is minimal.
}

In the first set of results, we locate the blast in
the Pacific about 180 kilometers off the coast of near Westport, Washington. 
This spot was chosen since it
is well studied by the earthquake-generated tsunami researchers due to
its proximity to the M9 Cascadia fault
\cite{petersenCramerFrankel:2002,noaa1}.
By the time
the waves reach shore they have decayed and are under a meter high. 
Since they do not
have the long length scales of earthquake tsunamis, we not see any
inundation on shore. 
In the second set of results 
we move the location offshore to Long Beach, California, where  
there is significant coastal infracture, and has also been studied
extensively in relation to earthquake tsunamis \cite{UsluEtAl:2010}.
We place the blast approximately 30 kilometers from shore, so
that there is less time for the waves to decay. 
In all simulations, bathymetry is available from the 
NOAA National Center for Environmental Information web site.

\begin{figure}[h!]
\begin{center}
\includegraphics[trim=2 10 10 15,clip,height=2.2in]{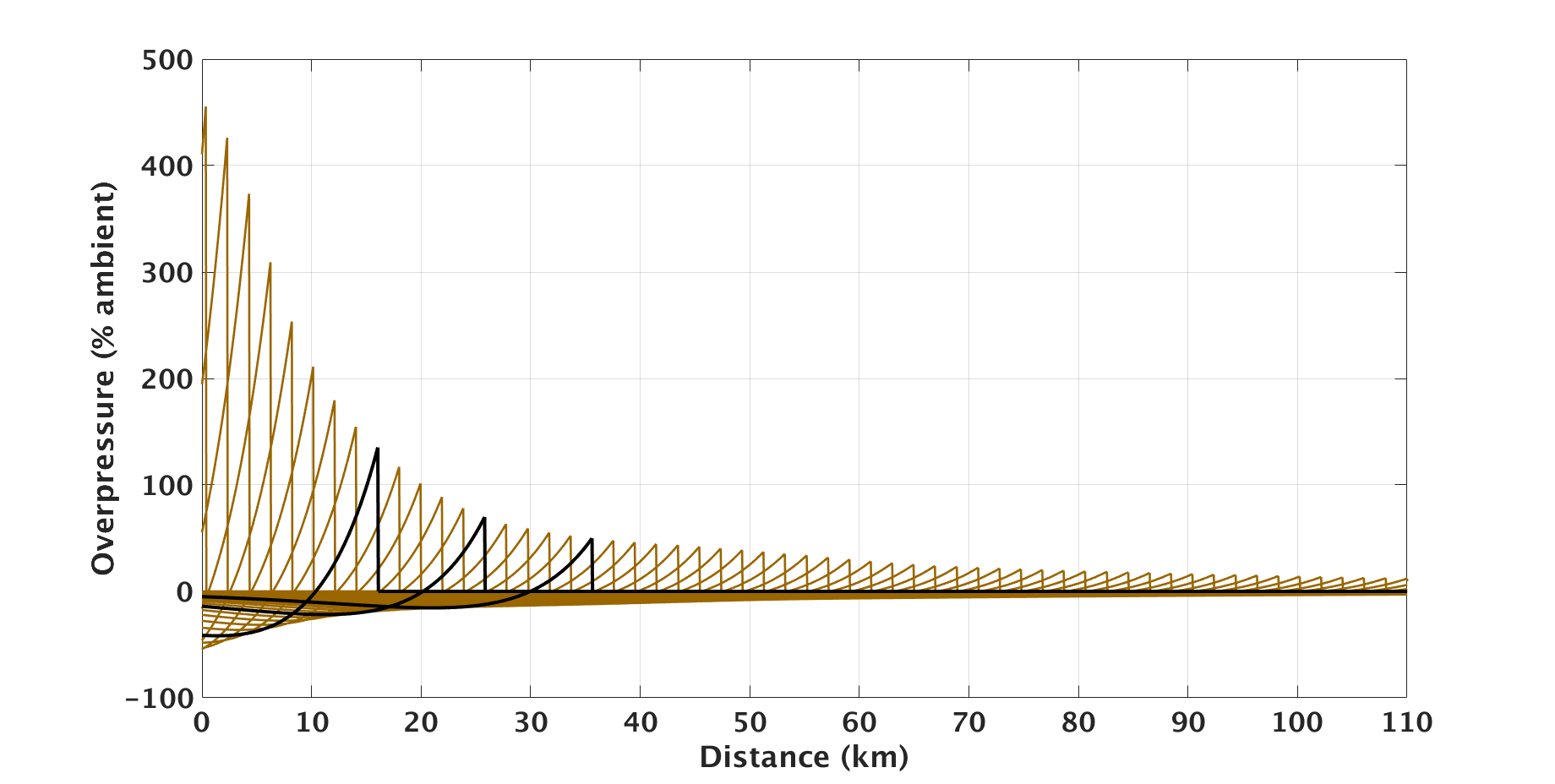}
\caption{\sf Ground footprint for 
250 MT blast wave overpressure as a function of distance
from the initial blast.  The curves are drawn every 5 seconds.  A few of the curves are drawn in black to more
clearly show a typical Friedlander profile. }
\label{fig:250model}
\end{center}
\end{figure}

To perform these simulations, we use a model of the blast wave simulated
in \cite{aftosmis:bolide2016}.  The ground footprint for the
overpressure was extracted, a Friedlander profile  was fit to the 
data, and its amplitude as a function of time  was modeled
by a sum of Gaussians. The 250MT model
is shown in Fig.~\ref{fig:250model}, with a few of the profiles 
drawn in black to illustrate their form. The profiles start with the
rise in pressure from the incoming blast, and are followed by the
expected rarefaction wave (underpressure) some distance behind.  Note
that the maximum amplitude is over 4 atmospheres, but decays rapidly 
from its initial peak. 
\commentout{
The 100MT blast has the same form, but the maximum amplitude of the
overpressure  is 2.3 atmospheres.
}
In the model, the blast wave travels at a fixed speed of 391 m/sec. 
This may be less accurate at early times. If the asteroid enters at a low angle of 
incidence, the blast wave travels more quickly 
when it first hits
the ground. This would also lead to a more anisotropic
response when see from the ground. 
Here, however, we assume the blast wave is radially symmetric. We then use
this model of the overpressure as
a source term in our two-dimensional shallow water simulations using
the software package {\tt GeoClaw}. 

{\tt GeoClaw} is an open source software package developed since 
1994 \cite{LeVequeGeorgeBerger:actaNumerica} for modeling 
geophysical flows with bathymetry using the shallow water equations.
It is mostly used for simulations of
tsunami generation, propagation and inundation. 
{\tt GeoClaw} uses a well-balanced, second-order finite volume scheme
for the numerics \cite{article:LeVeque97,George:2008}. 
Some of the  strengths of {\tt GeoClaw} include automatic
tracking of coastal inundation, robustness in its
handling of dry states, a local adaptive mesh
refinement capability, and the automated setup
that allows for multiple bathymetry input files with varying resolution.
A bottom friction term is included using a constant Manning 
coefficient of 0.025. 
The results below do not include a Coriolis force, which
we have found to be unimportant. There is no dispersion in the shallow
water equations.
In 2011 the code was approved by the U.S. National
Tsunami Hazard Mitigation Program (NTHMP) after an extensive set of
benchmarks used to verify and validate the code \cite{NTHMP}. 

\subsection{Westport Results}

\begin{figure}[b!]
\vspace*{-.2in}
\begin{center}
\includegraphics[trim=0 10 0 0,clip,width=.9\textwidth]{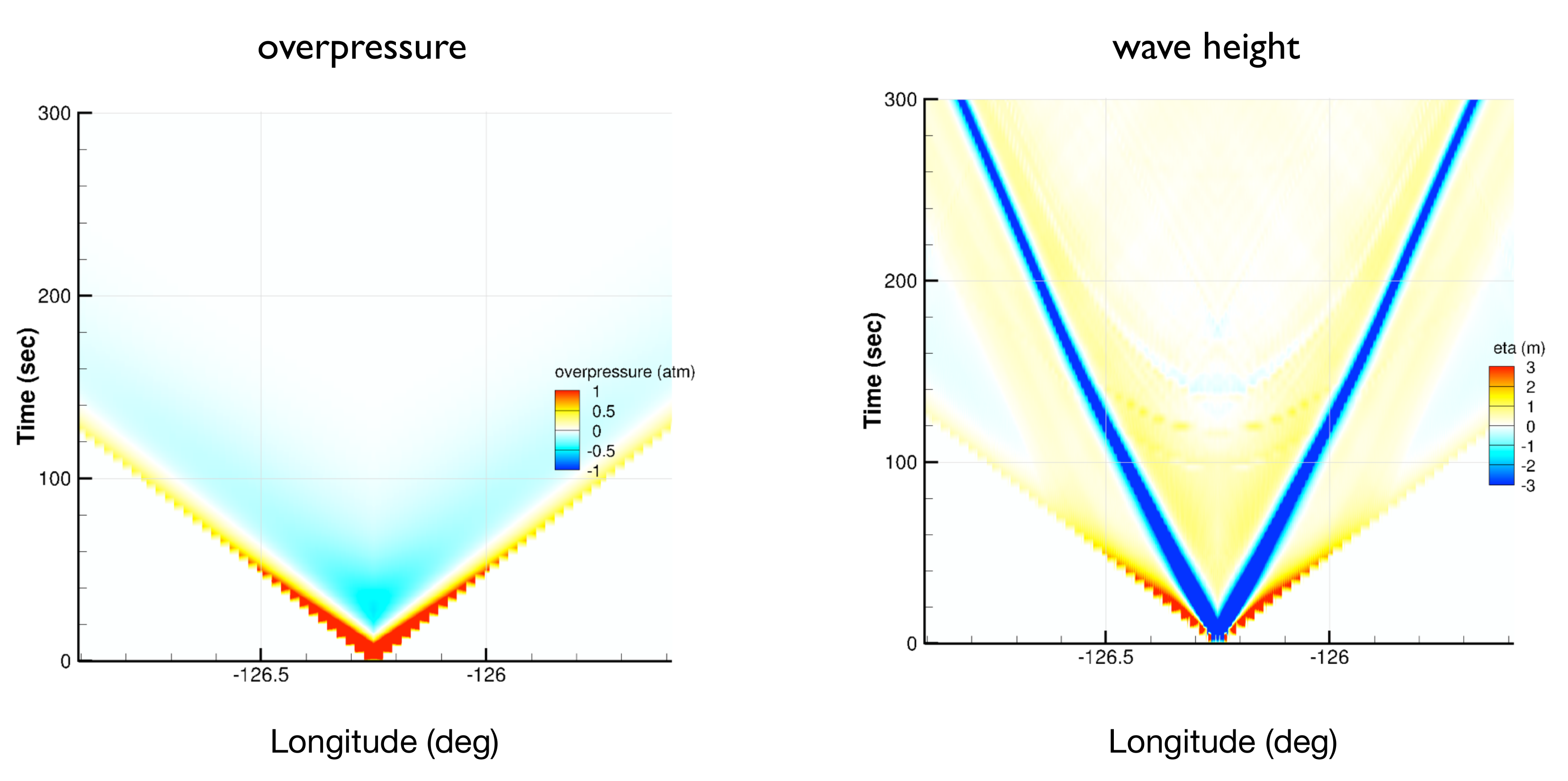}
\caption{\sf The Hovm{\"o}ller plot shows the 
overpressure in atmospheres through the center of the blast location 
(left)  and the wave height (right).
The blast wave speed is approximately twice the gravity wave speed, and its
amplitude decays more rapidly.}
\label{fig:westport_hov}
\end{center}
\vspace*{-.2in}
\end{figure}

For this set of simulations, the 250MT blast is located at 
$-126.25^{\circ}$ longitude and $46.99^{\circ}$ latitude, about 30
kilometers from the continental slope. The ocean
is 2575 meters deep at this spot.
The blast location is about 180 kilometers from shore.
Many simulations were performed with different  mesh resolutions.  
The finest grids used in the adaptive simulations
had a resolution of 1/3 arc second.
Three bathymetric data sets were used -- a 1 minute resolution
covering the whole domain, 
a 3 second resolution nearer shore, and a 1/3 arc second bathymetry
that included the shoreline itself. 

In Fig. \ref{fig:westport_hov} we show a Hovm{\"o}ller plot through the
center of the blast location at a fixed latitude. 
On the left is  the atmospheric overpressure for the first 300 seconds.
This is the forcing that travels at 391 m/sec. 
On the right is the amplitude of the
water's response. Two waves traveling at different speeds are visible.
A shallow water gravity wave travels with speed
$\sqrt{gh}$, which at the blast location is 158 m/sec.
It is evident  that the blast wave travels approximately twice as fast
as the gravity wave. The blast wave reaches
the edge of the graph in just over 150
seconds instead of the 300 seconds of the main water wave (in blue,
since it is a depression).  
Also visible in the wave height plot is a wave that starts off
in red and travels at the same speed as the blast wave, and whose
amplitude decays more rapidly.
Here the color scale
saturates below the maximuym the maximum value in ecah plot so that
smaller waves are visible.

\begin{figure}[h!]
\begin{center}
\includegraphics[trim=0 20 0 20,clip,width=.7\textwidth]{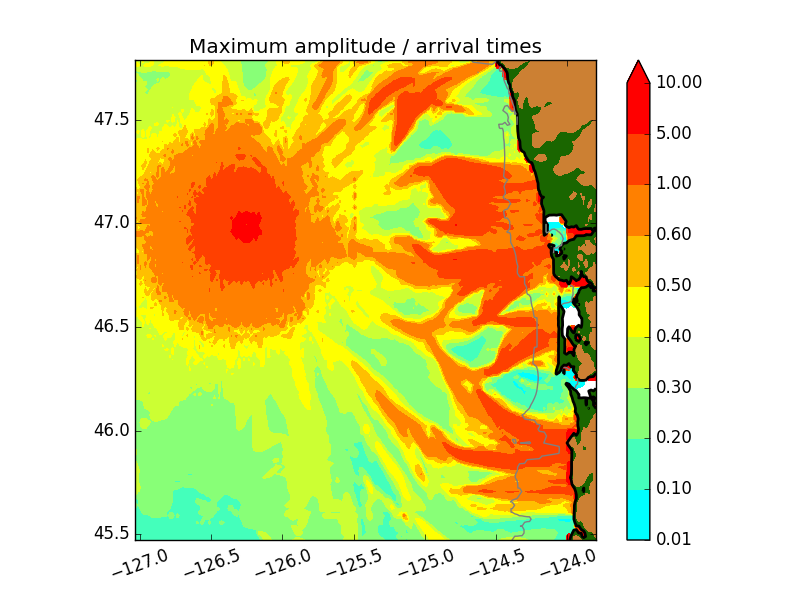}
\caption{\sf Maximum amplitude found between the blast location and the
shoreline during the simulation.} 
\label{fig:Pacific_maxAmp}
\end{center}
\vspace*{-.1in}
\end{figure}

Fig.~\ref{fig:Pacific_maxAmp} shows the maximum amplitude
found at any time in the simulation at that location. Note that
the color bar  is
not linear in this plot, so that the different levels can more easily
be seen. Nearest the blast location the maximum wave amplitude is over
10 meters, but it decay rapidly.  As the waves approach shore, the waves
are amplified in a non-uniform way by the bathymetry.
The coastline is outlined in black. The light gray contour line
represents the location of the waves after 30 minutes.
We do not see any inundation of land, although admittedly at this resolution it
would be hard to see.


\begin{figure}[h!]
\vspace*{-.1in}
\begin{center}
\includegraphics[trim=0 30 0 20,clip,height=2.2in]{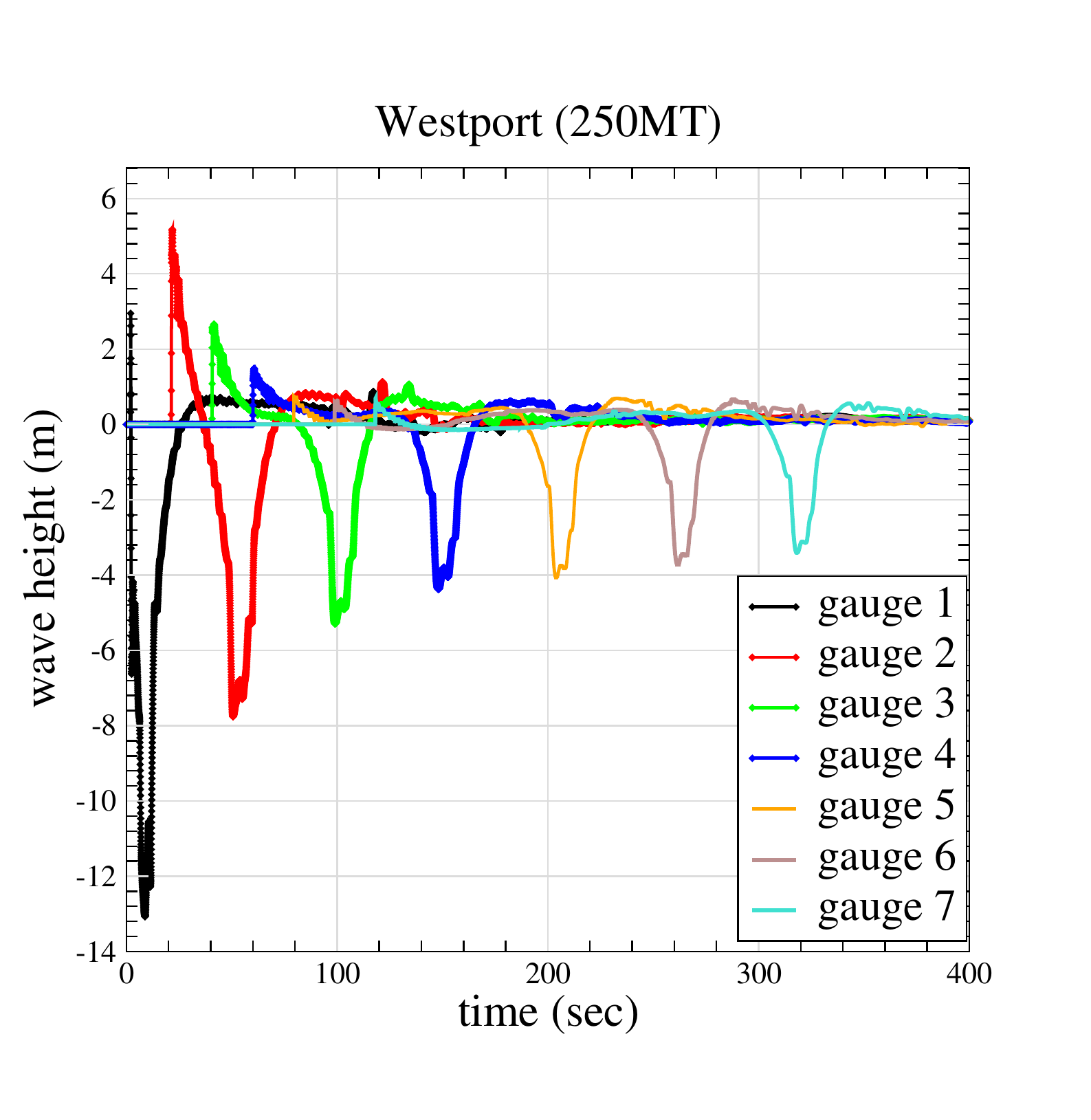}
\hspace*{.2in}
\includegraphics[trim=0 30 0 20,clip,height=2.2in]{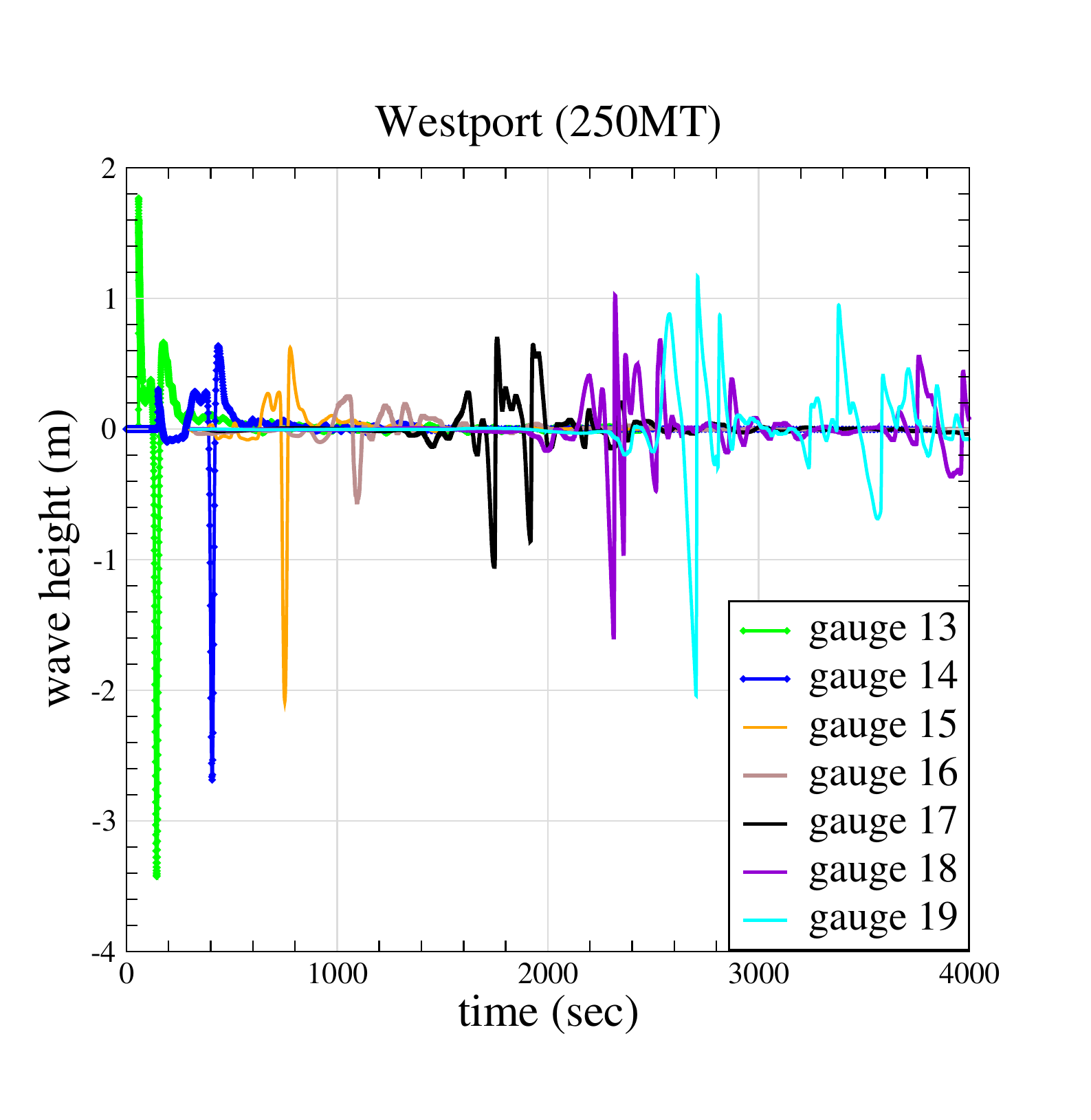}
\caption{\sf (Left) Gauges near blast location, every $ 0.1^\circ$
starting $0.01^\circ$ from blast. Gauges show rapid decay
of maximum amplitude, much slower decay of maximum depressions. 
(Right) Gauges approaching shoreline show
similar wave forms with decreasing maximum amplitude before shoaling 
increases it. These gauge locations are marked in
Fig.~\ref{fig:westport_zoom}.}
\label{fig:westport_gaugeResponses}
\end{center}
\vspace*{-.1in}
\end{figure}

\begin{figure}[h!]
\vspace*{.1in}
\begin{center}
\includegraphics[width=\textwidth]{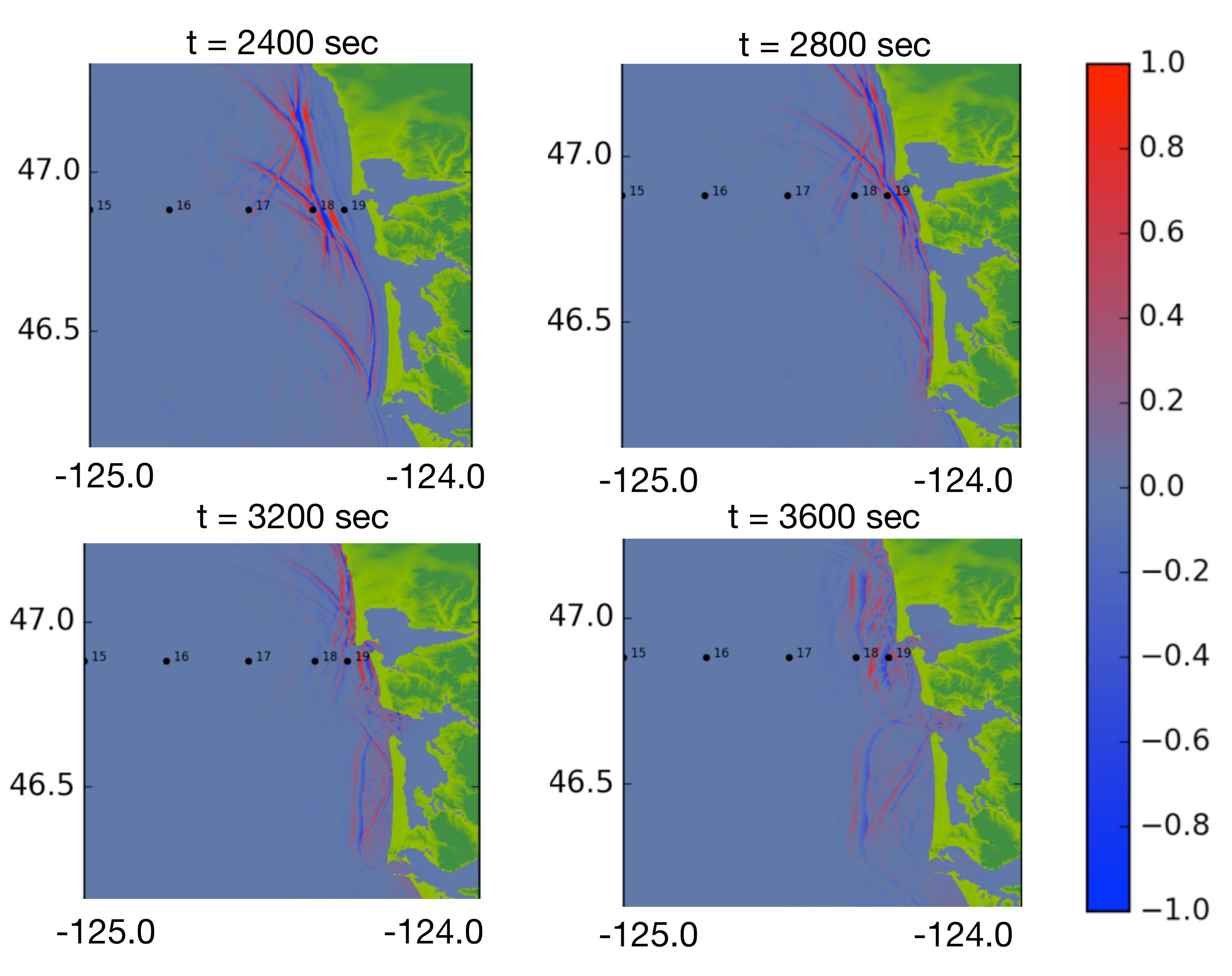}
\caption{\sf Zoom of waves approaching shoreline around Westport. No
inundation is observed.}
\label{fig:westport_zoom}
\end{center}
\vspace*{.1in}
\end{figure}

Fig.~\ref{fig:westport_gaugeResponses} shows
the time history of wave heights through several gauge plots.  
The left plot shows 7 gauges placed  $0.1^\circ$  apart (about 10 kilometers 
at this latitude), starting about 1 kilometer from the blast.  
The gauge closest to the blast location has a maximum amplitude that reaches 5
meters.  Subsequent gauges show a very rapid decay in
maximum amplitude. These positive elevation waves are the 
water's response to the blast wave overpressure, and travel at the same
speed as the blast wave.
Most of the ocean's response at this 
location appears as a depression, not an elevation. 
The negative amplitude  wave travels at the gravity wave speed, $\sqrt
{g h}$, where the water has depth $h$. It shows much less decay in
amplitude. For example, looking at gauge 3 and 5, the peak amplitude
decays from 2.7 meters to 0.72 meters in about 50 seconds, whereas 
between 100 and 200 seconds, the trough decays from -5.3 meters  
to -4.1 meters in about 100 seconds.

Fig.~\ref{fig:westport_gaugeResponses} right shows
gauges approaching the shore, starting about 100 kilometers away from
the blast. These are not equally spaced but are placed from .25 to
$.1^\circ$
apart (from 25 to 10 kilometers at this latitude),  becoming closer as they 
approach shore and the bathymetry  changes
more rapidly. Shoaling is observed as the wave amplitudes increase, 
seen in gauges 17 and higher.
The maximum elevation is between 0.5 and 1 meters, and 
its duration is short, at least compared to
earthquake-generated tsunamis.

Finally, Fig.~\ref{fig:westport_zoom} shows several close-ups of the region
near shore. The waves are of uneven strength due to focusing from the
bathymetry. The maximum amplitude is around 1 meter. The sequence shows
waves reflecting from the coastline but not flooding it.  Some waves enter
Grays Harbor, but they small amplitude and do not flood the
inland area either.  

\subsection{Long Beach Results}

\begin{figure}[h!]
\vspace*{-.1in}
\begin{center}
\includegraphics[trim=0 0 0 12,clip,width=.8\textwidth]{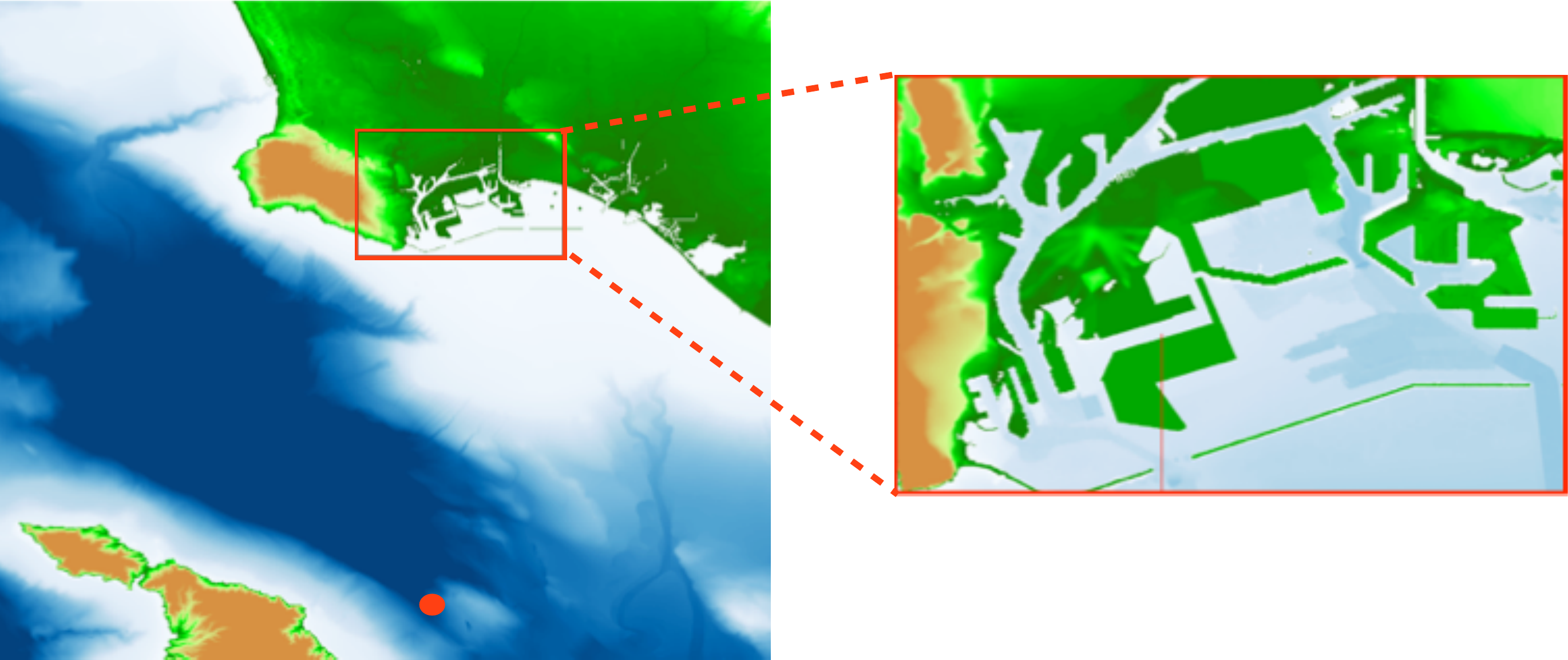}
\caption{\sf Location of air burst northeast of Catalina, and zoom of Long Beach shoreline.} 
\label{fig:longBeachTopo}
\end{center}
\vspace*{-.1in}
\end{figure}

For this set of experiments we move the simulations to
Long Beach, California. 
We locate the blast very close to shore so that the waves do not
have time to decay. 
Again we have detailed bathymetry at a resolution of 1/3 arc
second between Catalina Island and Long Beach, and use a 1 minute
dataset outside of this region.  The blast is 
located at $-118.25^\circ$ longitude and $33.41^\circ$ latitude, where
the ocean is 797 meters deep.  This is about 30 kilometers from shore.
Fig.~\ref{fig:longBeachTopo} shows the region where  the blast is located,
and a zoom of the Long Beach harbor where we will look for flooding.

\begin{figure}[h!]
\vspace*{-.1in}
\begin{center}
\makebox[\linewidth][c]{
\includegraphics[width=\textwidth]{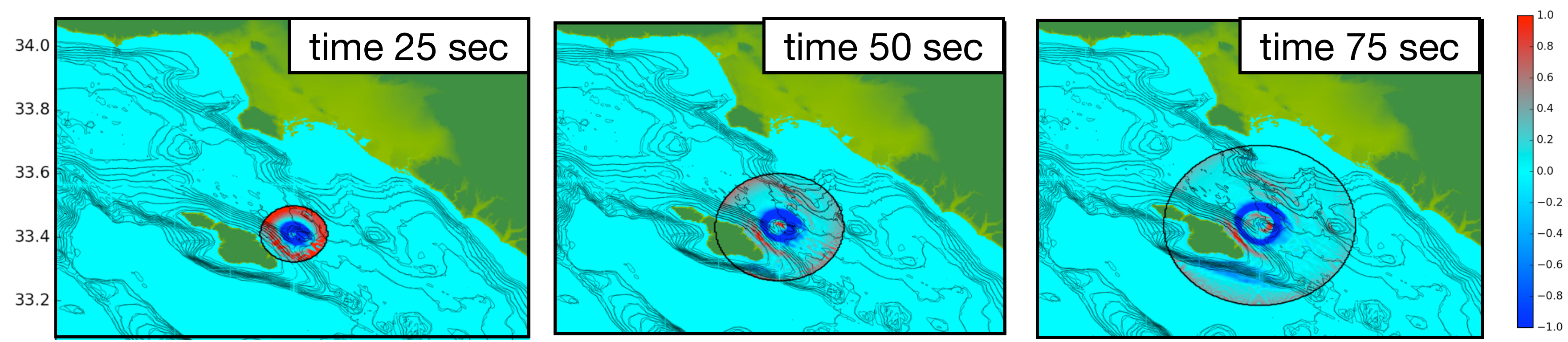}
}
\caption{\sf Snapshots at early times of blast wave and ocean waves in Long
Beach simulation.} 
\label{fig:LB_snapshots}
\end{center}
\vspace*{-.1in}
\end{figure}

Fig.~\ref{fig:LB_snapshots} shows the ocean response at several points in
time.  The black circle on the plots indicates the location of the
air blast. In the plot marked at 25 seconds, note how the wave height in
red that is closest to the blast location is not as circular as the air
blast itself.
It also decays faster than in the Westport computation.
This will be explained by the model problem presented in the next section.

There is a breakwater that protects long Beach. It reflects 
most of the waves that reach it, with only a
small portion getting through the opening.  Waves
that reach the harbor go around the breakwater, and are reflected from the
shoreline back into this region.

\begin{figure}[h!]
\vspace*{-.18in}
\begin{center}
\includegraphics[trim=50 100 0 40,clip,width=.758\textwidth]{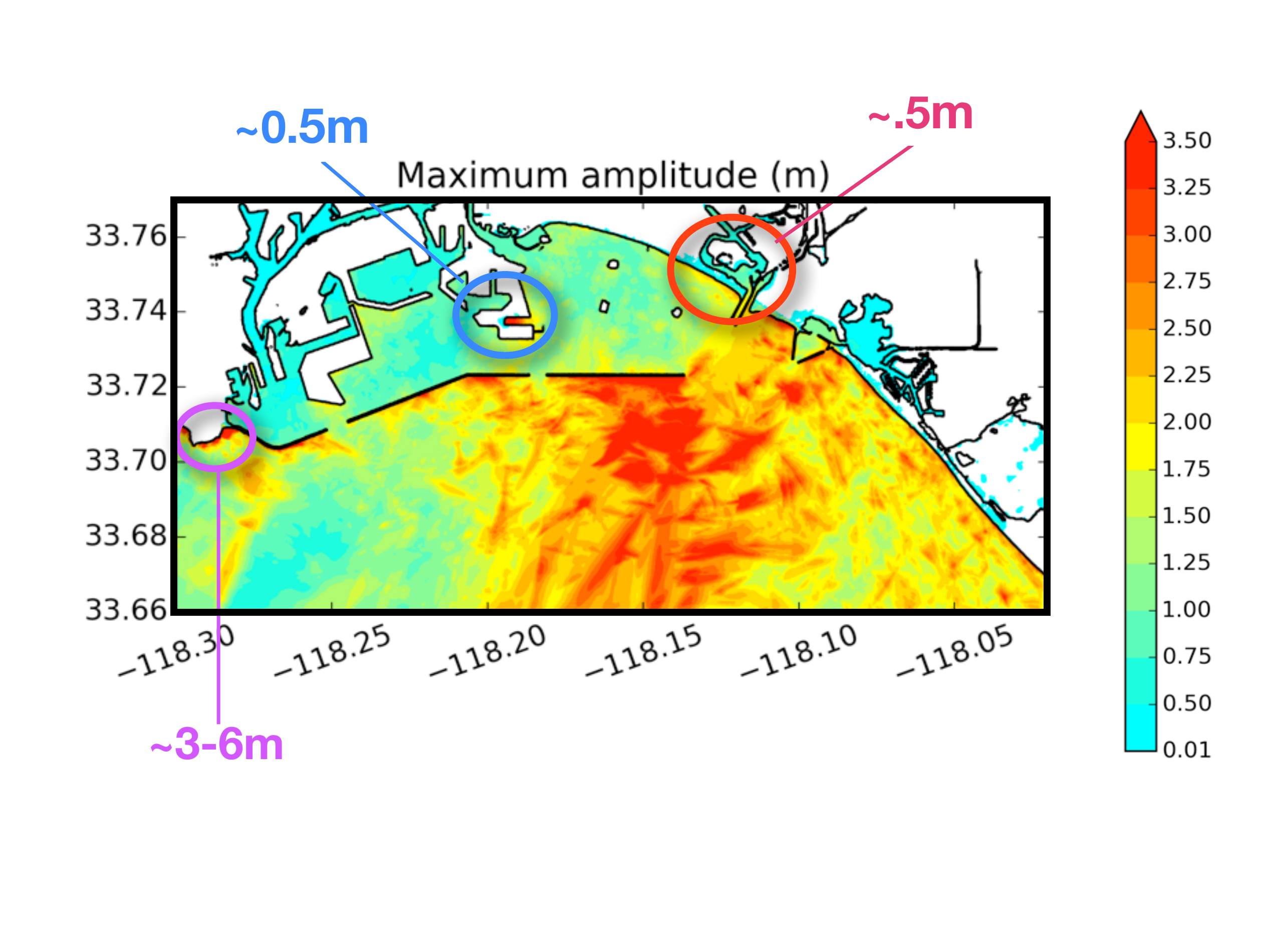}
\caption{\sf Maximum amplitude plot shows 0.5 meters of water overtopping 
the dock and the riverbank.}
\label{fig:LB_maxArr}
\end{center}
\end{figure}

Fig.~\ref{fig:LB_maxArr} shows a plot of the maximum water amplitudes
seen in the harbor area.  We do see some overtopping of land, but it is
very small. In several locations it reaches 0.5 meters, where the inlet
exceeds its boundaries, and on the dock in the middle. The region with
the largest 
accumulation is just outside the harbor before the breakwater, 
where the maximum amplitude seen is
between 3 and 6 meters. There is a steep cliff here however and the
water does not propagate inland. Paradoxically, in other experiments
where the blast was located closer to shore by a factor of 2, there was no
overtopping.  This can also be 
explained by our model problem in the next section.

\section{Shallow water model} \label{sec:sw}

In this section we present a one-dimensional model of the shallow water
equations (SWE)
that explains much of the behavior seen in the previous
examples. In the SWE, the atmospheric overpressure appears as an
external forcing $p_e$ in the momentum equation. In one space dimension
it is
\begin{equation}
\begin{split}
h_t + (hu)_x  =&\, 0 \\
(hu)_t + (hu^2 + \onehalf g h^2)_x =&\,  \frac{-h \, {p_e}_x}{\rho_w},
\end{split}
\label{eqn:1dsw}
\end{equation}
where $h$ is the  height of the water surface over the bottom, $u$
is the depth-averaged velocity of the water in the $x$ direction, and
$g$ is gravity, and $\rho_w$ is the density of water.
${p_e}$ is the external pressure forcing, and it's $x$ derivative 
is ${p_e}_x$.
We assume constant bathymetry in the model. 
See \cite{vreugdenhil} for these equations, or \cite{mandli:thesis} for a
complete derivation.
In this section (and the next), 
the conclusions are in the last few paragraphs after the analysis.

\subsection{Derivation and Analysis}
As stated in the introduction, we simplify the pressure forcing by
assuming it has the form of a traveling wave, and look for solutions
$h$ and $u$ that are traveling waves too.
This means they are functions only of the moving variable
\[
        m = x-st \; .
\]
so that $\partial_t h = -s \partial_m h(x-st) = -sh_m$.
The equations (\ref{eqn:1dsw}) become a pair of 
ordinary differential equations 
\begin{align}
\label{eqn:mvar1}
-s h_m + (hu)_m &= 0 \\
-s (hu)_m + (hu^2 + \onehalf g h^2)_m &= \frac{\displaystyle -h {p_e}_m}{\displaystyle \rho_w} \;.
\label{eqn:mvar2}
\end{align}

Equation \eqref{eqn:mvar1} can be integrated to give 
$-s h + h u = \mbox{\em const}$. 
We evaluate the constant by taking $m \to\infty$, 
where $u \to 0$ and $h \to h_0$, with $h_0$ the undisturbed water height.
(We assume the overpressure has localized support, and goes to zero as
$ m \rightarrow \infty$.)
Therefore $-s h + hu = -s h_0$.
This may be re-written as 
\[
        u(m) = \frac{s(h(m) - h_0 )}{h(m)} \; .
\]
We use this to eliminate $u$ from \eqref{eqn:mvar2}, which gives 
\begin{equation*}
-s \, (s(h-h_0))_m  + \left(\frac{s^2(h-h_0)^2}{h}\right)_m + \left(\onehalf g h^2\right)_m =
            \frac{-h {p_e}_m}{\rho_w} \; .
\end{equation*}
After some algebra, this leads to
\begin{equation}
\frac{s^2}{2} \, \left(\frac{h_0^2}{h^2}\right)_m  + g \,  \pr{h} = \frac{ -{p_e}_m}{\rho_w} \;.
\label{eqn:mvar2Only}
\end{equation}
As before, this may be integrated exactly.
Again we use the boundary conditions $h \to h_0$, $u \to 0$, and $p_e \to 0$ as $m \to \infty$.
The result is
\begin{equation}
   \frac{s^2 }{2} \left(1 - \frac{h_0^2}{h(m)^2} \right) + g h_0 \left(1 - \frac{h(m)}{h_0}
   \right) = \frac{ p_e(m)}{\rho_w} \; .
\label{eqn:res2} 
\end{equation}

To summarize, eq.~\eqref{eqn:res2} is the water's
response according to shallow water theory.
The solution of the differential equation system (\ref{eqn:mvar1}) and (\ref{eqn:mvar2})
is an algebraic relation between the overpressure and the response height.
It tells us that the water height at a point $m=x-st$ is determined by the 
overpressure at the same point.

To get a better feel for the behavior of the solution \eqref{eqn:res2}
we linearize it, writing
$h(m) = h_0 + h_r(m)$ where $h_r(m)$ is the response height.
The linearization uses the relation
\[
\left(\frac{h_0}{h(m)}\right)^2 =  \frac{h_0^2}{\left(h_0 + h_r\right)^2} \approx 1 - \frac{2h_r}{h_0} \; ,
\]
which is valid when $h_r \ll h_0$. This is our case, since the change in wave height
$h_r$ is a number in meters
where  $h_0$ is typically measured in kilometers.
The linearization of \eqref{eqn:res2} is:
\begin{equation}
h_r = \frac{h_0 \, p_e}{\rho_w (s^2 - c_w^2)} \; .
\label{eqn:res2lin}
\end{equation}

\begin{figure}
\begin{center}
\includegraphics[height=2.2in,trim=0 0 0 30,clip]{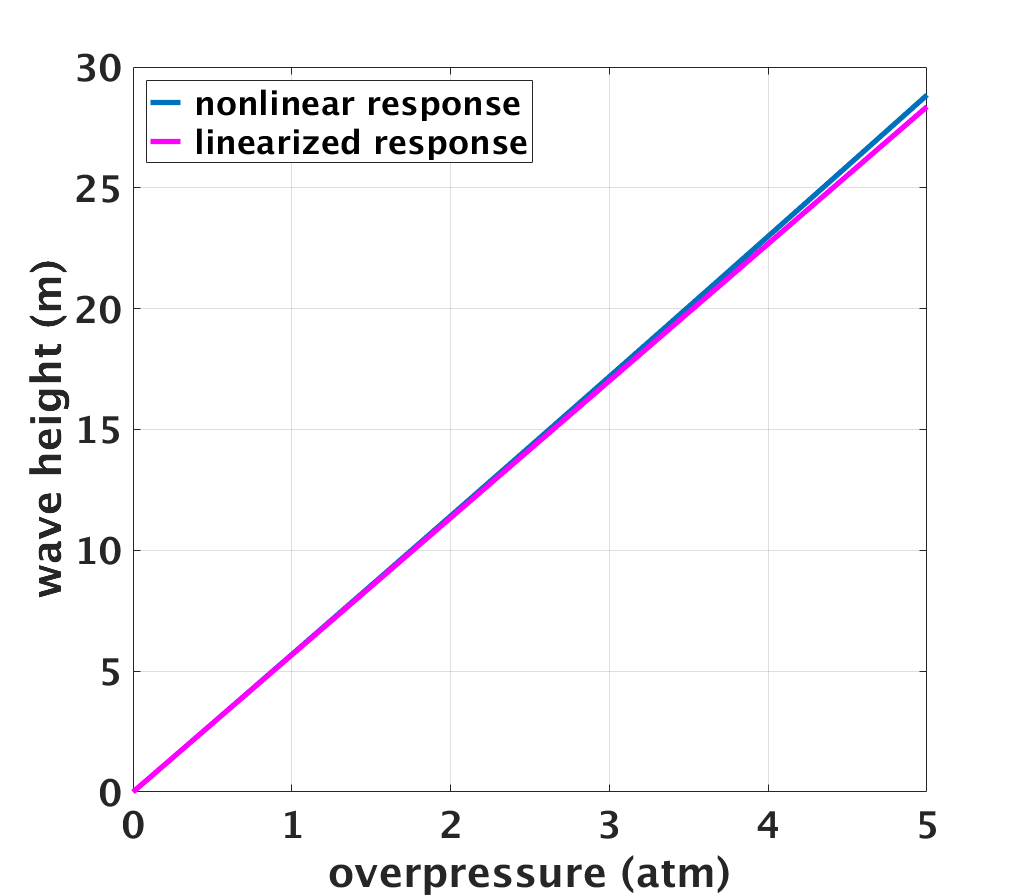}
\includegraphics[height=2.2in,trim=0 0 0 30,clip]{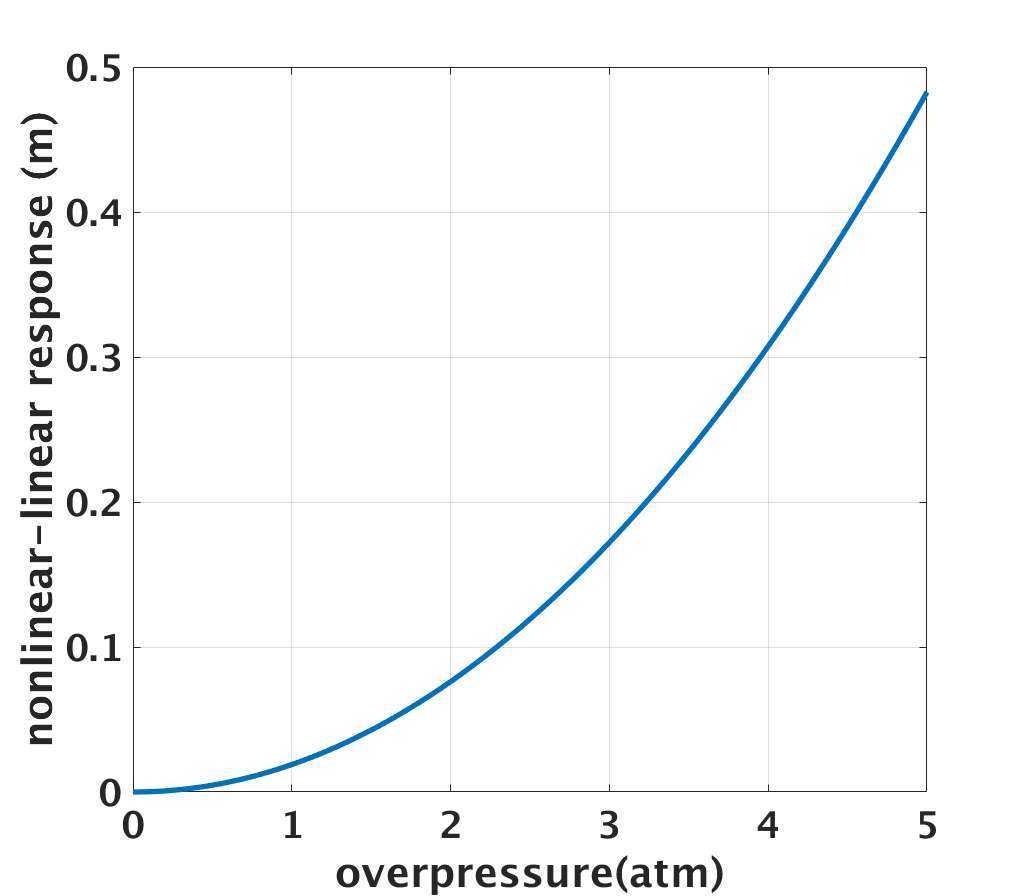}
\caption{\sf Wave height as a function of overpressure, for the nonlinear equation 
\eqref{eqn:res2} and the linearized equation \eqref{eqn:res2lin},
using $h_0 = 4$km and $s = 350$ m/sec.  
The curves are very close. The
right figure is a plot of their the difference, which is on the order of
a percent.}
\label{fig:swe}
\end{center}
\end{figure}

Fig.~\ref{fig:swe} shows that the full response theory \eqref{eqn:res2} and the linear 
approximation \eqref{eqn:res2lin} are very close to each other for the parameters of interest.
The plot uses a constant depth of $h_0=4$~km, and takes 
$\rho_w = 1025\,\mbox{kg}/\mbox{m}^3$.
The maximum difference between the
nonlinear and linear wave heights in Fig.~\ref{fig:swe}
is half a meter, when the overpressure is five atmospheres.

To enumerate the consequences of the response predicted 
by \eqref{eqn:res2} and \eqref{eqn:res2lin}, we observe:
\begin{enumerate}
\item The response wave height $h_r$ is linearly proportional to the depth $h_0$.
A pressure wave over deep ocean has a stronger effect than a pressure
wave over a shallower continental shelf.

This explains why locating the blast in the Long Beach case 
closer to shore had less of an
effect.  If the distance offshore of the air blast 
from Long Beach is halved to 15
kilometers, the ocean is only
90 meter deep, resulting in approximately 1/10 the impact response.  
On the other hand, there is almost no difference in the decay rate of
the shallow water waves before they reach shore. 

\item If $s > c_w$, then $p_e$ and $h_r$ have the same sign.
The response height is positive in regions of positive overpressure.
This contradicts an intuition that positive overpressure would depress the water surface.
This response is similar to the case of a forced oscillator in
vibrational analysis.
Consider for example $\ddot{x}= -x + A \cos(\omega t)$.
The steady solution is 
\[
       x(t) = \frac{A}{1 - \omega^2} \,\cos(\omega t) \; .
\]
For $\omega > 1$, the response $x(t)$ has the opposite sign from the forcing $A\cos(\omega t)$.
For pressure forcings with speeds slower than the water speed, the water
response would be a depression, with $h_r$ negative.

This response is clearly seen in the all the simulations. The wave that
travels at the speed of the blast wave is an elevation.  However, in the Long
Beach results, we can see that the response wave is not
uniformly circular when the depth of the water changes rapidly.
Note that since the speed of sound in air is 343 m/sec, the water would
have to be more than 12 kilometers deep for the gravity wave speed to
exceed the speed of the pressure forcing.  Hence in all cases on earth
we expect an elevation of the sea surface beneath the pressure wave.

\item The response is particularly strong when the forcing speed $s$ is close to the 
gravity wave speed $c_w \approx 200$ m/sec., (for $h_0 = 4 $ km).
In this case we have a Proudman resonance
\cite{Proudman:1929,MonserratVilibicRabinovich:2006}.
This is the regime for meteo-tsunamis, in basins whose depth 
leads to gravity wave speeds that match the squall speeds. 
These speeds are much slower than the speed of sound in air.
\end{enumerate}

\subsection{Shallow Water Model Computations}\label{sec:SWECompRes}
This subsection illustrates the behavior of  the analytic model 
described above with numerical simulations. 
We solve
the equation set  \eqref{eqn:1dsw} again using {\tt GeoClaw},
with $u = h = p_e = 0$ 
for $t \le  0$, and 
$$
p_e = p_{ambient}  \exp(-0.1(x-st)^2) ,
$$
with $p_{ambient}$ = 1 atm.  for $t>0$.  

These initial conditions use an impulsive start for the air blast
pressure wave at time $t=0.0$, so it is not a traveling wave.  
This generates gravity waves, one 
moving left and the
other moving right, at speeds $ c = \pm \sqrt{g h_0}$, in addition to the 
forced water
wave traveling at speed $s$.  
The exact solution to the linearized shallow water equations can be
found by matching the conditions at time 0.
It is a combination of the forced wave solution to the inhomogeneous
equation, presented above in equation \eqref{eqn:res2lin}, plus the 
solution to the homogeneous equation. Writing the full solution in terms
of the response wave $h_r$ gives:
\begin{equation}
h(x,t) = h_r(x-st) - \left(\frac{s}{c} + 1\right) \frac{h_r(x-ct)}{2}
 + \left(\frac{s}{c} - 1\right) \frac{h_r(x+ct)}{2},
\label{eqn:fullLinRes}
\end{equation}
consisting of a left-going and right-going gravity wave traveling with
speed $c$, along with what we have been calling the response wave.

Eq.~ \eqref{eqn:fullLinRes} shows that the left-going tsunami 
wave will have a smaller amplitude in absolute
value for $s/c>1$ than the right going wave. Also, the latter
will be a depression, since it has amplitude $-0.5 \cdot (1+s/c))$.

\begin{figure}
\begin{center}
\includegraphics[height=2.4in,trim=0 20 0 0,clip]{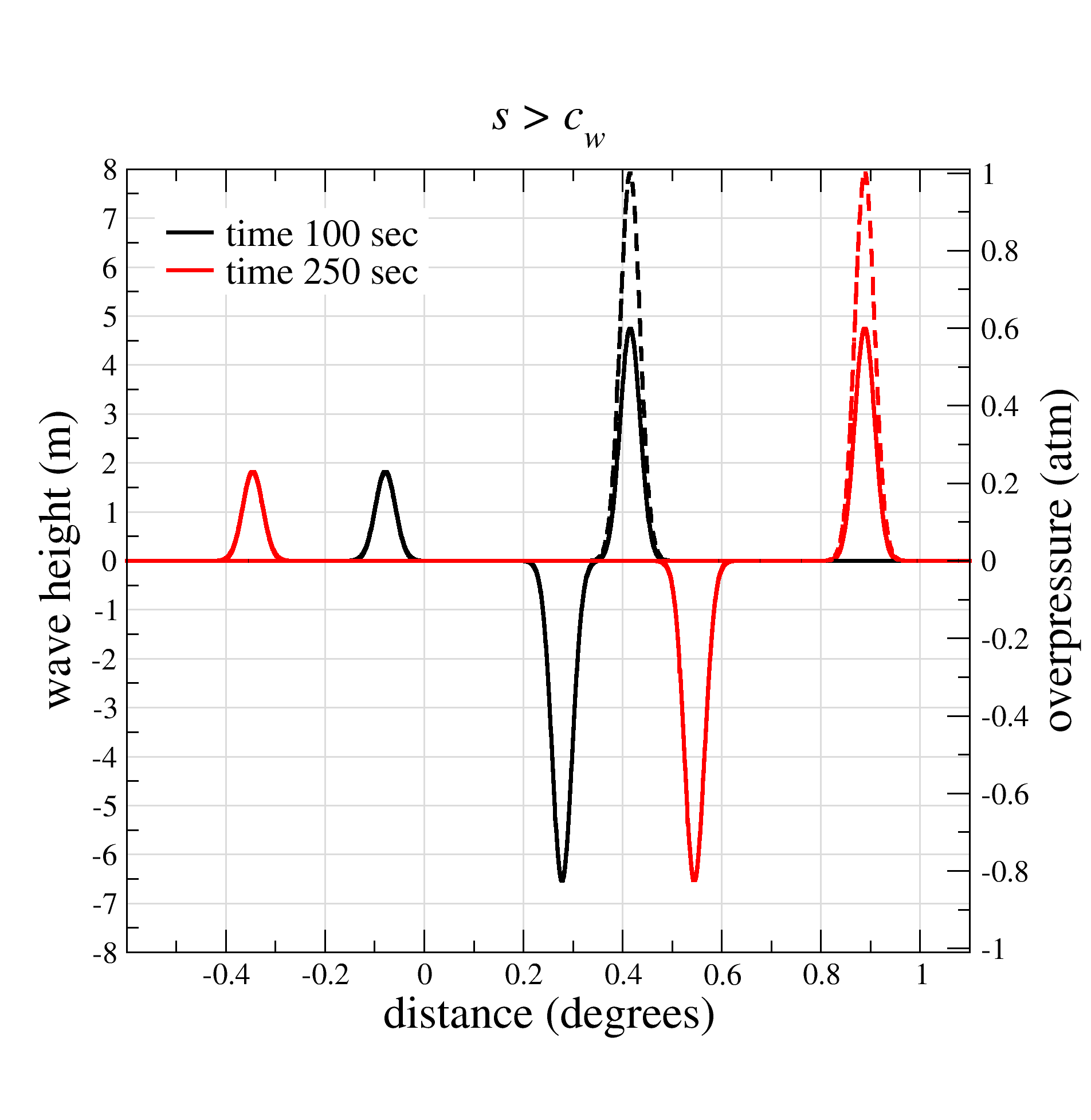}
\hspace*{.3in}
\includegraphics[height=2.4in,trim=0 20 0 0,clip]{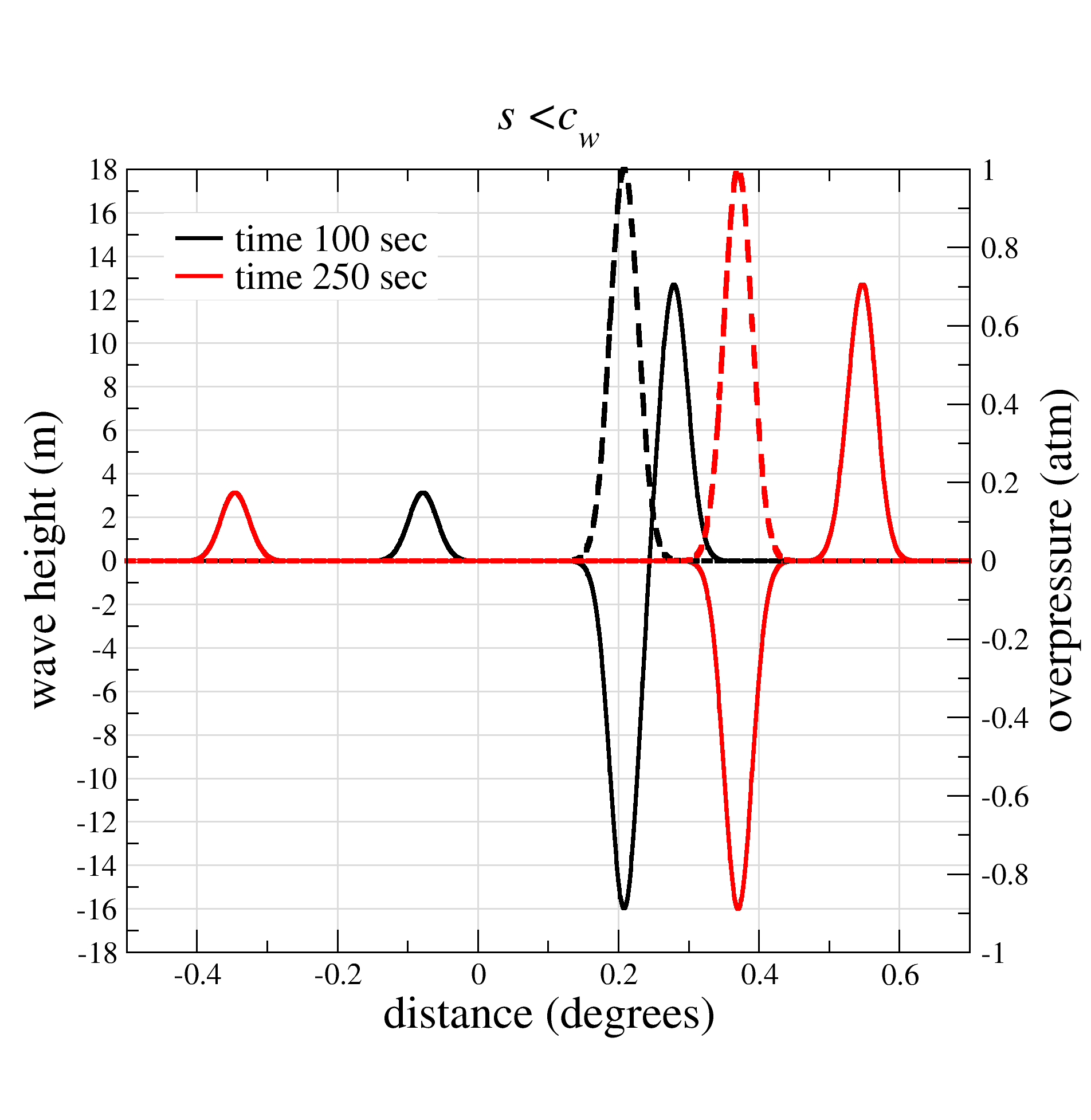}
\caption{\sf Numerical simulations showing wave heights $h_r$ and overpressure
$p_e$ at two 
different times. Left experiment uses $s > c_w$, on the right the pressure front is slower. 
On the left, the  wave heights (solid line)  
under the pressure pulse (dashed line) are positive, as equation \eqref{eqn:res2lin}
predicts, and the tsunami wave trails the pressure wave. On the right, the pressure pulse is
above a negative wave height (depression), and the tsunami wave leads the pressure wave.} 
\label{fig:res1}
\end{center}
\end{figure}

Numerical results illustrating this are shown in Fig.~\ref{fig:res1}.
The figure contains two curves for each time. Solid lines show the water
wave heights, and dashed lines show the air overpressure profiles.
In Fig.~\ref{fig:res1} left, the speed $s = .350$~km/sec, somewhat larger than the speed of
sound in air. 
For this case, since $s>c_w$, (for $h_0$ = 4 km, $c_w = \sqrt{g h_0} \simeq$  0.198~km/sec), 
the forced wave height is positive since the overpressure is.
Note that the gravity wave at the same point in time trails the pressure wave.
The right-moving gravity wave is a depression,
the left moving wave is a smaller elevation.  
Since this calculation is in one space dimension, the waves
do not decay.  In the two dimensional shallow water equations, the gravity wave decays with
the square root of distance.  Also, the pressure blast wave, and therefore the leading
water response would both decay too.

By contrast, Fig.~\ref{fig:res1} right shows the water's response
for an overpressure moving at $0.120$~km/sec, slower than the gravity
wave ($s < c_w$).
The tsunami waves travel at the same speed in both computations,  but they 
have different amplitudes and signs. The tsunami wave is the opposite sign as the wave due
to the pressure. This is consistent with 
conservation of mass.

\begin{figure}
\begin{center}
\textbf{\hspace*{-.15in}Ramped Gaussian pressure profile \hspace*{.8in}  Friedlander pressure profile}\par\medskip
\includegraphics[width=.45\textwidth,trim=70 30 70 20,clip]{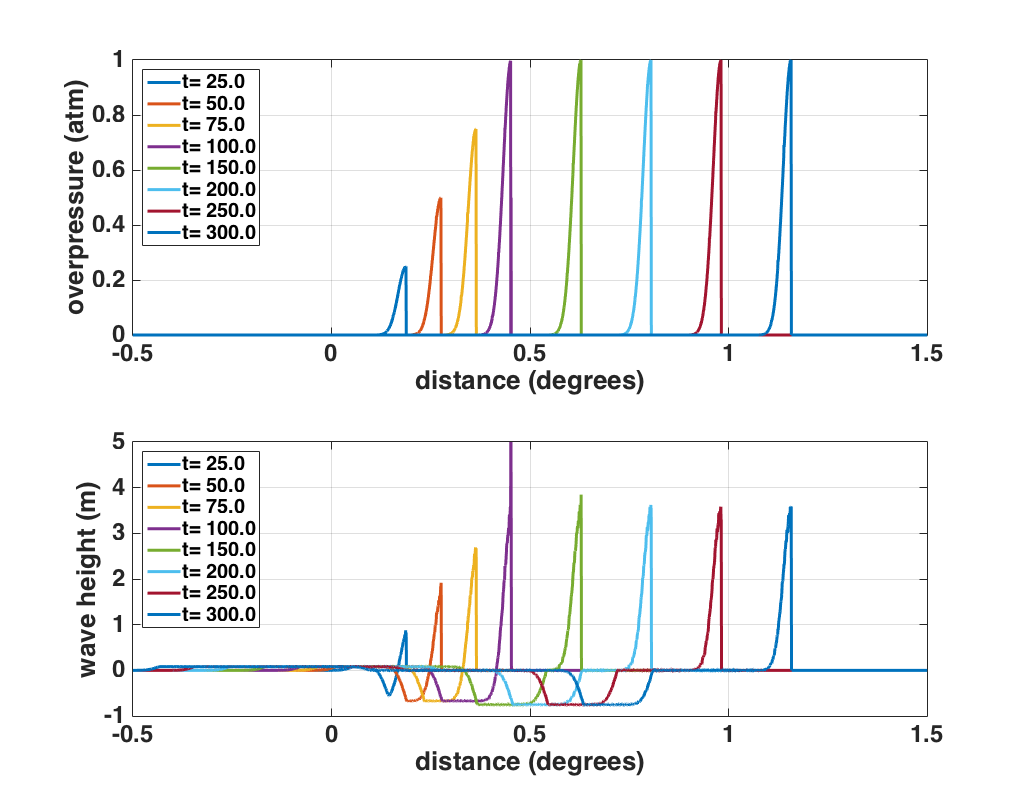}
\hspace*{.4in}
\includegraphics[width=.45\textwidth,trim=50 30 80 20,clip]{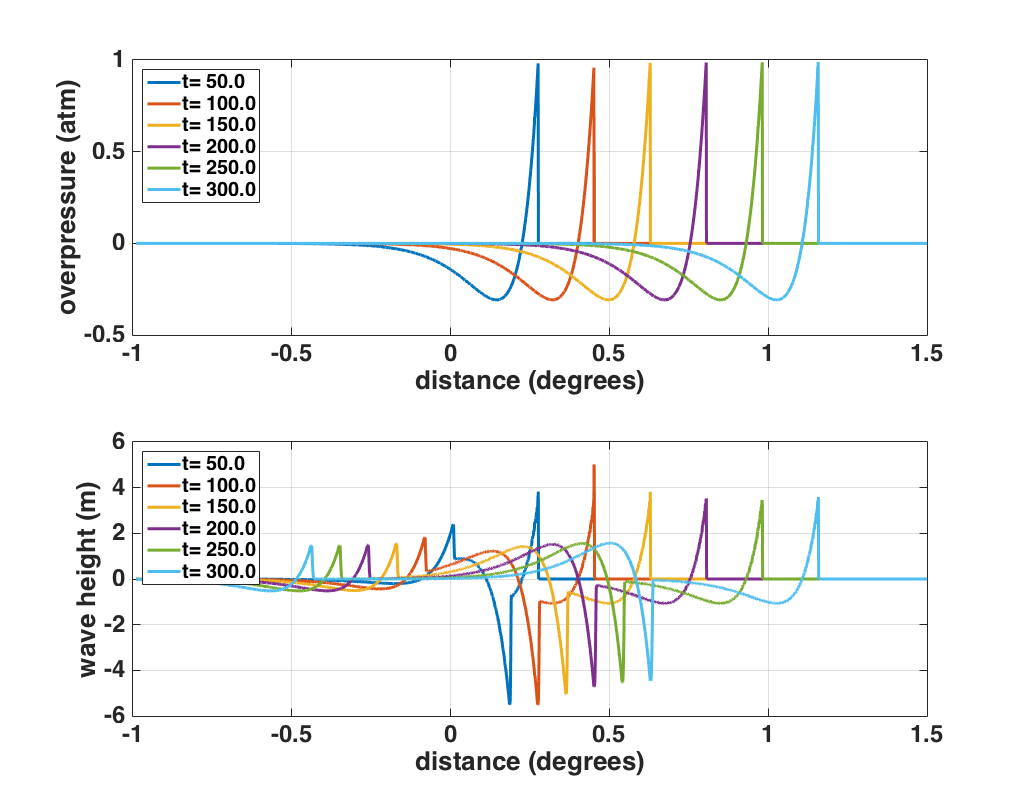}
\caption{\sf Left figure uses same high-speed Gaussian pressure pulse
as in Fig.~\ref{fig:res1}  but with the amplitude
linearly ramped up over 100 seconds. Right figures uses a
Friedlander blast wave profile 
instead of a Gaussian.
Both figures show the same positive forced water wave (since $s>c_w$)  
and the expected gravity waves from \eqref{eqn:fullLinRes}. 
Note that different scales are used in the two plots.}
\label{fig:diffForcing}
\end{center}
\end{figure}

To give a more complete picture, two more experiments with $s > c_w$ but different
forcings are shown in Figure \ref{fig:diffForcing}.
The figures on the
left use a Gaussian pressure forcing but their magnitude is ramped up for
the first 100 seconds. This results in quite a different-looking gravity
wave.   
The right figure uses a typical Friedlander
blast profile described in section \ref{sec:sims} for the overpressure,
but keeping the amplitude constant at 1 atm. It looks similar to
the  Gaussian example above.

\section{Linearized Euler model} \label{sec:le}

In this section we analyze a more complete model
of the ocean's response to an airburst, to uncover possible 
shortcomings of the shallow water 
model of Section \ref{sec:sw}.
We model the water using the Euler equations of a compressible fluid,
which will bring in the effects of compressibility and dispersion.
Another possibility would be to use one of the forms of the
Boussinesq equations, but
that also assumes incompressible flow, and would be more difficult to
analyze. (See however a nice comparison of 
SWE and Serre-Green-Nagdhi Boussinesq results in \cite{Popinet:2015}.)
We continue to neglect Coriolis forces, viscosity, friction,  the Earth's curvature, etc.
We linearize the Euler equations and the boundary conditions, since 
Fig.~\ref{fig:swe} suggests that linear approximations are reasonably accurate for
these parameters. 

\subsection{Derivation and Analysis}

Our starting point for this section is the linearized Euler equations
with linearized boundary conditions.  A derivation 
is given in the Appendix.
An explicit solution is not possible, and the results will 
depend instead on wave number. 
We will use wave number 
$k = \frac{2 \pi}{L}$, where the length scale $L$
for the atmospheric pressure wave is on the order of $10 - 20$
kilometers. This is very short relative to  earthquake-generated
tsunamis, which can have length scales on the order of 100 kilometers or
more.  
As before, those not interested in the analysis can skip to the end of
the section for a summary of the main points.

The linearized Euler  equations and boundary conditions
that we use for analysis are
\begin{equation}
\begin{split}
\widetilde{\rho}_t + \rho_w \widetilde{u}_x + \rho_w \widetilde{w}_z &= 0 \\[3pt]
\rho_w \widetilde{u}_t + c_a^2 \widetilde{\rho}_x &= 0 \\[3pt]
\rho_w \widetilde{w}_t + c_a^2 \widetilde{\rho}_z &= -\widetilde{\rho} g \; ,
\end{split}
\label{eqn:linCompEulerRepeat}
\end{equation}
where $c_a$ is the speed of sound in water.
(We use $c_a$ for acoustic to distinguish it from the gravity wave speed
$c_w = \sqrt{gh}$).
Here, $\widetilde{\rho}$ is a small perturbation of $\rho$ (and the same
for the other variables), except for
\[
     h(x,t) = h_0 + h_r(x,t) \; .
\]
where $h_r$ is again the water's disturbance height for consistency with the
previous section.
The boundary conditions are:
\begin{align}
\text{bottom:} \hspace*{1.15in} \widetilde{w}(x,z=0,t) &= 0 \hspace*{1.5in}  \\
\text{top:} \hspace*{1.6in} \frac{\partial h_r (x,t) }{\partial t} &= \widetilde{w}(x,h_0,t) \\
\text{pressure bc:}\quad c_a^2\widetilde{\rho}(x,h_0,t)- \rho_w  g \, {h_r}(x,t) &= p_e (x,t) .
\label{eqn:pbc}
\end{align}

As in section \ref{sec:sw}, we will assume the atmospheric pressure forcing has the form 
$p_e(x-st)$, and look for solutions of the same form, functions of $m = x-st$ and $z$.  
The system \eqref{eqn:linCompEulerRepeat} becomes
\begin{subequations}
\begin{align}
-s \widetilde{\rho}_m + \rho_w \widetilde{u}_m + \rho_w \widetilde{w}_z &= 0 
        \label{eqn:rho_m} \\[3pt]
-s \rho_w \widetilde{u}_m + c_a^2 \widetilde{\rho}_m &= 0 
        \label{eqn:u_m}\\[3pt]
-s \rho_w \widetilde{w}_m + c_a^2 \widetilde{\rho}_z &= -\widetilde{\rho} g \; .
\label{eqn:w_m}
\end{align}
\end{subequations}
The boundary conditions become
\begin{subequations}
\begin{align}
\widetilde{w}(m,0) &= 0   \label{eqn:eulermbc1} \\
\widetilde{w}(m,h_0) &= -s \, h_{r,m}(m)  \label{eqn:eulermbc2} \\
c_a^2 \widetilde{\rho}(m,h_0) &= \rho_w g h_r(m) + p_e(m) \; .  \label{eqn:eulermbc3}
\end{align}
\end{subequations}
This system now includes 
the effects of dispersion and water compressibility.

These equations cannot be solved in closed form for general $p_e$.
Therefore, we study the response using Fourier analysis.
We will take a Fourier mode of the overpressure 
\begin{equation}
        p_e(m) = A_k e^{ikm} \; ,
\label{eqn:pressuremode}
\end{equation}
with amplitude $A_k$ and compute the response as a function of $m$.
The responses will have the form
\begin{subequations}
\begin{align}
h_r(m) &= \,  \widehat{h}_r \,\, e^{ikm} \label{eqn:vardefa} \\
\widetilde{\rho}(m,z) &= \widehat{\rho}(z) e^{ikm} \label{eqn:vardefb} \\
\widetilde{u}(m,z) &= \widehat{u}(z) e^{ikm} \label{eqn:vardefc} \\
\widetilde{w}(m,z) &= \widehat{w}(z) e^{ikm} \; .  \label{eqn:vardefd}
\end{align}
\end{subequations}
The hat variables are the Fourier multipliers.
The partial differential equations (\ref{eqn:rho_m}-c) become 
ordinary differential equations 
with wave number $k$ as a parameter.


Note that \eqref{eqn:u_m} depends only on derivatives with respect to $m$.
Integrating it gives
\begin{equation*}
-s \rho_w \ut + c_a^2 \rhot = 0.
\end{equation*}
The constant of integration is zero for each $z$
since as $m \rightarrow\infty$ we know $\ut=0$ and $\rhot = 0$.
This gives an expression for $\rhot$ in terms of $\ut$,
\begin{equation}
\rhot = \frac{s \rho_w \ut}{c_a^2},
\label{eqn:rhotDef}
\end{equation}
which we can use in \eqref{eqn:rho_m} and \eqref{eqn:w_m}. 
After substituting for $\rhot$ and dividing by $\rho_w$, the remaining 
system of two equations is
\begin{equation}
\begin{aligned}
\ut_m \left(1-\frac{s^2}{c_a^2}\right) + \wt_z &= 0 \\
-\wt_m \quad + \ut_z &= -\frac{g}{c_a^2} \ut.
\label{eqn:uwOnly}
\end{aligned}
\end{equation}

Substituting the Fourier modes  (\ref{eqn:vardefc}-d) into \eqref{eqn:uwOnly}, and differentiating $\ut$ and $\wt$ 
with respect to $m$ gives an ordinary  differential equation in $z$ for the velocities,
\begin{equation}
\begin{pmatrix}  \uh \\ \wh  \end{pmatrix}_z = 
\begin{pmatrix} 
 - g/c_a^2 \,  \uh + i \, k \, \wh \\
-i\,  k \,  \uh \, (1-s^2/c_a^2) 
\end{pmatrix} = 
\begin{bmatrix}  - g/c_a^2 & i \, k  \\  -i\,  k \, (1-s^2/c_a^2) & 0   \end{bmatrix}
\begin{pmatrix}  \uh \\ \wh  \end{pmatrix}  \; .
\label{eqn:ode}
\end{equation}
The general solution to this 2-by-2 system is the linear combination 
\begin{equation}
\begin{pmatrix}  
\uh \\ \wh  \end{pmatrix}  = 
a_+ {\bf v_+} e^{\mu_+ z} + a_- {\bf v_-} e^{\mu_- z}  \; ,
\label{eqn:gensoln}
\end{equation}
where $\mu_{\pm}$ and {$\bf v_{\pm}$}
are the eigenvalues and eigenvectors of the matrix in \eqref{eqn:ode},
and the scalar coefficients $a_{\pm}$ are chosen to satisfy the boundary
conditions.
The eigenvalues are 
\begin{equation}
    \mu_{\pm} =  \frac{ \frac{-g}{c_a^2} \pm \sqrt{ \frac{g^2}{c_a^4} + 4
    k^2 (1-s^2/c_a^2) }}{2} \; .
\label{eqn:mu}
\end{equation}
The eigenvectors (chosen to make the algebra easier so they are not normalized) are
\begin{equation}
{\bf
v_+ = \begin{pmatrix}  
\cfrac{2 \mu_+}{-i k } \\[9pt]
2 (1-s^2/c_a^2)  
\end{pmatrix} \; ,
\hspace{.7in}
v_- = \begin{pmatrix}  
\cfrac{2 \mu_-}{-i k } \\[9pt]
2 (1-s^2/c_a^2)   
\end{pmatrix}
} \; .
\label{eqn:evec}
\end{equation}

The boundary condition at $z=0$ is \eqref{eqn:eulermbc1}. To apply it, note that
$\wh$ corresponds to the second component of the eigenvectors
${\bf v}_{\pm}$.
We find that $a_+ = - a_-$.
Henceforth we call this coefficient simply $a$.

Next, we substitute the Fourier modes  (\ref{eqn:vardefc}-d) into the remaining boundary conditions
\eqref{eqn:eulermbc2} and \eqref{eqn:eulermbc3}.
We use the pressure forcing equation \eqref{eqn:pressuremode} in the form $\widehat{p}_e =
A_k$.
The result is
\begin{subequations}
\begin{align}
\wh(h_0)  = - i  k   s  \,  \widehat{h}_r \label{eqn:bchata}\\
c_a^2 \rhoh(h_0) - \rho_w g \widehat{h}_r = A_k \; .
\label{eqn:bchatb}
\end{align}
\end{subequations}
Using equation \eqref{eqn:rhotDef} to substitute $ \rhoh = \frac{s
\rho_w}{c_a^2} \uh$  in \eqref{eqn:bchatb} gives an expression for $\widehat{h}_r$
\begin{equation}
\widehat{h}_r = \frac{s}{g} \uh - \frac{A_k}{\rho_w g} \; .
\label{eqn:waveht}
\end{equation}
This can be used to replace $\widehat{h}_r$ in \eqref{eqn:bchata} to get 
\begin{equation}
\wh(h_0) = \frac{- i k s^2}{g}  \uh(h_0) + \frac{iksA_k}{\rho_w g} \; .
\label{eqn:lastEqn}
\end{equation}
The final steps are using the form of the solution \eqref{eqn:gensoln} in \eqref{eqn:lastEqn}
to solve for the coefficient $a$.  With this, everything is known, 
and $\uh$, $\wh$ and the
response height $\widehat{h}_r$ can be evaluated.

Putting it all together we get
\begin{equation}
 2 \, a (1-s^2/c_a^2) \left(  e^{\mu_+ h_0} - e^{\mu_- h_0} \right) =
\frac{i k s^2}{g} \frac{2 \, a}{-i k} \, \left( \mu_+ e^{\mu_+ h_0} - \mu_- e^{\mu_- h_0} \right)
+ \frac{i k  s A_k}{\rho_w g} .
\label{eqn:prefinalMess}
\end{equation}
Grouping terms, the final expression to solve for $a$  (using
the definition \eqref{eqn:mu} for $\mu_{\pm}$) is given by
\begin{equation}
2 a \left[   (1-s^2/c_a^2) \left(  e^{\mu_+ h_0} - e^{\mu_- h_0} \right) +
\frac{ s^2}{ g} \, \left( \mu_+ e^{\mu_+ h_0} -
\mu_- e^{\mu_- h_0} \right) \right]
= \frac{i k  s A_k}{\rho_w g}
\label{eqn:finalMess}
\end{equation}

To summarize, given an overpressure amplitude $A_k$ with wavelength $k$, 
equation \eqref{eqn:finalMess} gives the scalar coefficient $a$ in the velocity
equations, then we solve for
$\widehat{u}$ and $\widehat{w}$ using \eqref{eqn:gensoln}, and use
\eqref{eqn:waveht} to get the Fourier multiplier for the wave height response.

\subsection{Linearized Euler Model Computations}

We evaluate these results using the following
parameters: an ocean with depth $h_0 = 4\,\mbox{km}$, 
ocean sound speed $c_a = 1500$~m/sec,
$\rho_w = 1025\,\mbox{kg/m}^3$,  
and  atmospheric overpressure of $A_k = 1 \,\mbox{atm}$ with
pressure wave speed $s=350$~m/sec, faster than the gravity
wave speed of about $200$~m/sec.
The responses are linear in the overpressure amplitude $A_k$, 
so we do not evaluate these curves 
for any other overpressures.

\begin{figure}
\vspace*{-.1in}
\begin{center}
\includegraphics[height=2.1in,trim=8 20 30 12,clip]{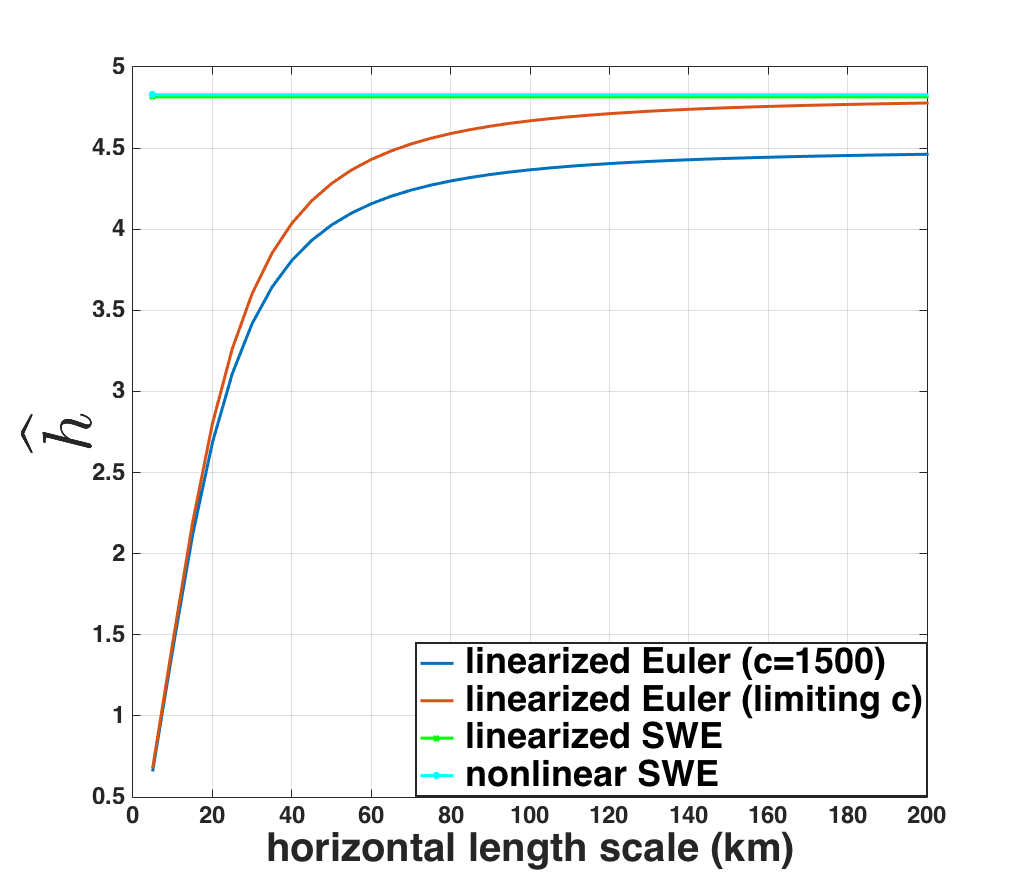}
\hspace*{.2in}
\includegraphics[height=2.1in,trim=10 20 30 12,clip]{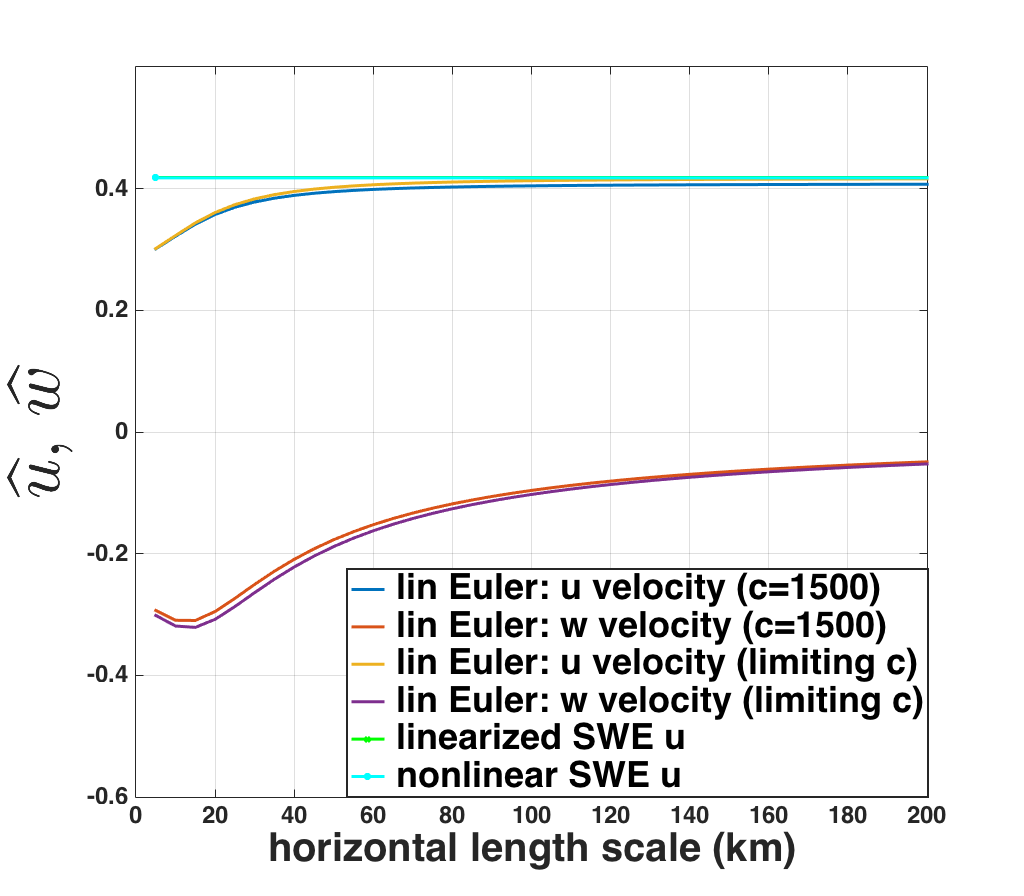}
\caption{\sf Left figure shows wave height $\widehat{h}(k)$
as a function of wavelength for the linearized
Euler equations using an atmospheric
overpressure of 1 atmosphere. Also shown is the 
shallow water solution from section \ref{sec:sw}. Right figure shows
the u and w velocities. Both plots show curves using the physical sound speed of
$c_a=1500$, and the limiting {\em infinite} speed solution. Both figures use the parameters
$h_0 = 4 $ km, and 1 atmosphere overpressure.}
\label{fig:euler}
\end{center}
\vspace*{-.1in}
\end{figure}

\begin{figure}[h]
\vspace*{-.1in}
\begin{center}
\includegraphics[height=2.2in,trim=0 5 0 44,clip]{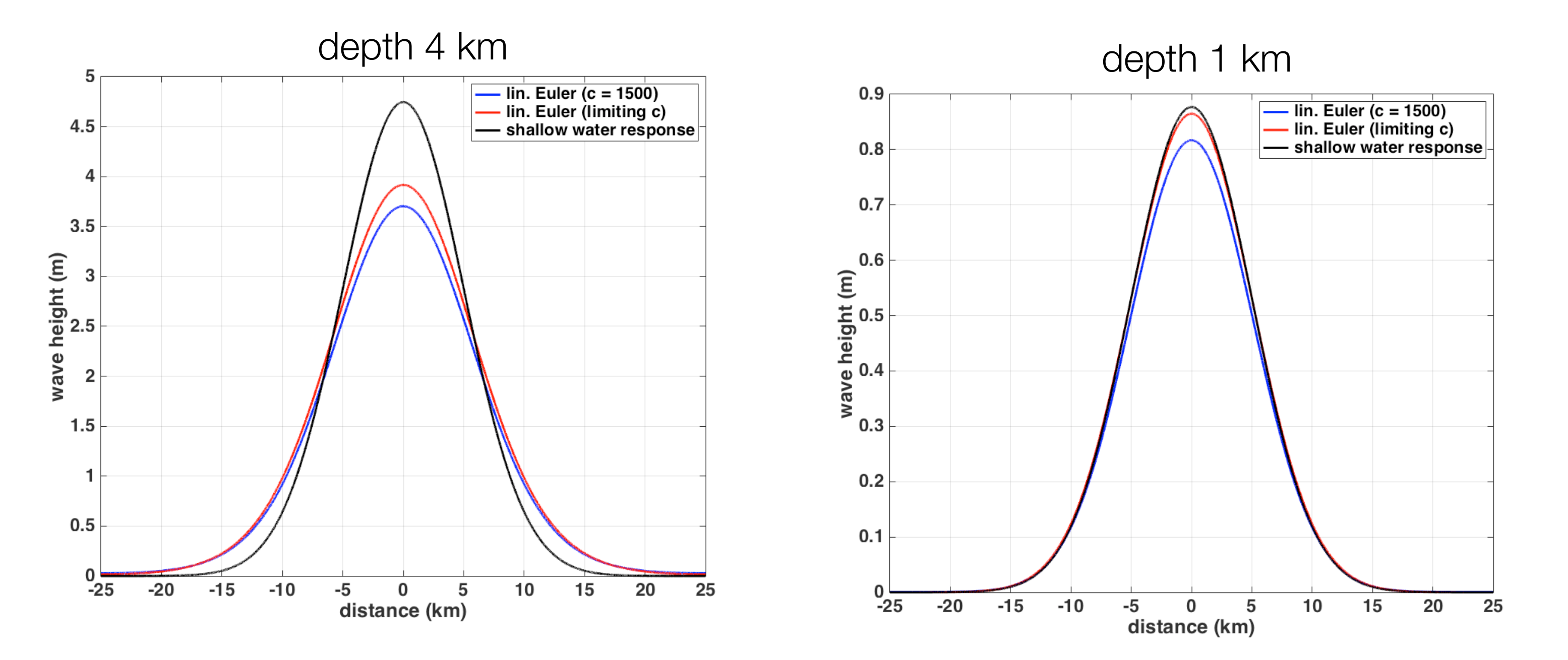}
\caption{Response to a Gaussian pressure pulse for the
linearized Euler equations, using the actual sound speed $c=1500$
m/sec, and a limiting sound speed  that mimics the incompressible case. The
shallow water response is also shown. Left uses depth $h_0$ = 4 km;
right uses $h_0$ = 1 km, so is closer to a shallow water wave. }
\label{fig:3gauss}
\end{center}
\vspace*{-.1in}
\end{figure}

Fig.~\ref{fig:euler} (left)  shows the surface wave height $\widehat{h}(k)$ 
as a function of length scale $L$, and 
(right) the amplitude of the surface velocities $\uh (h_0,k)$ and $\wh
(h_0,k)$ are shown.
There are two curves in each plot: one uses the physical acoustic 
water wave speed of $c_a = 1500$, and the other uses a very large non-physical 
acoustic speed in the water of $c_a \times 10^8$.
The latter corresponds to the intermediate model of finite depth but incompressible water.
This should, and does, asymptote in the long wave ($k \rightarrow 0$) 
limit to the result of the shallow water equations.
The difference between the blue and green curves
shows approximately a 20\% reduction in the amplitude of the longer
length scales due to compressibility (but that this is not the 
amplitude of the total wave response yet). 
Note also that  the $u$ velocity asymptotes to the shallow water limit,
and the $w$ velocity approaches zero. The velocity curves show less of an
effect due to compressibility.

For atmospheric forcing from asteroids with air bursts, the length scales 
of interest are closer to the short end,
perhaps 10 or 20 kilometers.
In this regime, the compressibility effects
are around 10\% or less. But at these wavelengths, dispersive effects
reduce the response predicted by shallow water theory by nearly half!

This becomes more clear  by comparing the forced wave response to
a Gaussian pressure pulse instead of using just a single  frequency.
We use the pressure pulse 
$$p_e(x-st) = p_{ambient} e^{-0.1  (x-st)^2},$$
take the Fourier transform, multiply by the Fourier multipliers shown in 
Fig.~\ref{fig:euler}, and transform back.
Figure \ref{fig:3gauss} shows the results for two different water
depths $h_0$: 4km and 1km. The blue curve uses the water
wave speed $c_a$=1500 m/sec, and the red curve uses the limiting $c_a$.
Compressibility changes the height by less than 10\% in both figures. 
However, in the deeper water, the shallow water response is 
almost 70\% larger, and has a narrower width since there is no
dispersion.  In the right figure, the water is shallower, and the
linearized Euler results are closer to the shallow water results. 

\begin{figure}[h!]
\begin{center}
\vspace*{-.1in}
\includegraphics[height=2.4in]{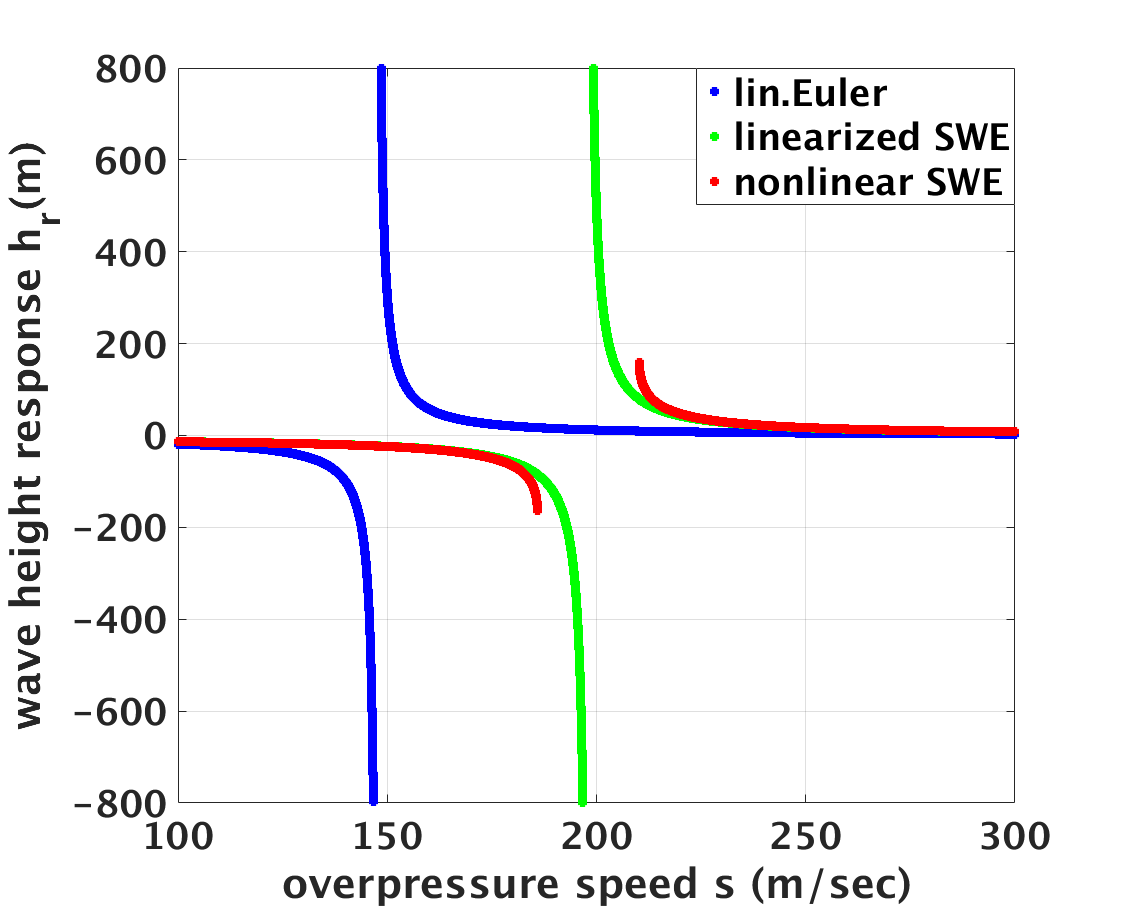}
\caption{\sf Wave height response as a function of  $s$,
the speed of the overpressure
front.  The depth $h_0$ is a constant 4 km, and the length scale is held 
fixed at 15 km.
There is a large variation between the models, especially in the
location where resonances occur.}
\label{fig:multiS}
\end{center}
\vspace*{-.1in}
\end{figure}

In Fig.~\ref{fig:multiS} we fix the horizontal length scale at 15 km
and instead vary the
speed of the pressure wave $s$. This figure again uses $h_0 = 4000$
meters, and $c_a = 1500$ m/sec.
Three curves are shown: the linearized Euler, and the nonlinear and linearized 
shallow water responses. There is much more difference in this set of curves,
particularly around the regions where resonance occurs.
Here too we see that the wave height response to the linearized 
Euler forcing is negative for pressure forcing speeds $s \lesssim 150$ and
again unintuitively, positive for larger $s$. 
There is also a  section of the red curve that is missing,
corrresponding to the regions where there is no smooth solution.
Note also that the overpressure speed where the resonance occurs
is significantly slower  for the linearized Euler
than for the SWE.

\section{Conclusions}   \label{sec:conc}

We have presented several numerical simulations using the shallow water
equations over real bathymetry that demonstrate the ocean's response to
a 250MT air-burst.  
There is no significant wave response from the ocean, in either the
forced wave or the gravity waves after a short distance. 
Our calculations show  that the amplitude of the pressure wave
response decreases much more rapidly than the gravity waves do. The
blast had to be very close to shore to get a sizeable response.
Thus the more serious danger from an air burst is not from the tsunami, 
but from the local effects of the blast wave itself. 

Several unexpected features found in the simulations
were explained using a one-dimensional model problem with a 
traveling wave solution for the SWE. 
One of the main results, that the wave response height  is proportional to the 
depth of the water, explains why putting the blast on a continental shelf 
close to shore did not generate more inundation than putting it 
further away in deeper water.

We also looked at the water's response to an air burst  using
the linearized Euler equations.
In this case the traveling wave model problem shows that the 
amplitudes of the important wave
numbers in the ocean's response  are greatly decreased. 
We do not yet know what this means for the gravity wave
response.
In addition, we expect the character of the water's
response to be different, since dispersive waves will generate 
a wave train characterized by multiple peaks and troughs.
The effect of this on land, and whether it causes 
inundation when the SWE response does not,  
is something we plan to investigate in the future.


\section{Appendix} \label{sec:app}

In this appendix we start with the nonlinear Euler equations for a
compressible inviscid fluid with nonlinear boundary conditions at the
interface between ocean and air.
The static unforced solution to these equations is determined by
hydrostatic balance. The hydrostatic pressure is $p_0$, and the
hydrostatic density is $\rho_0$.  
Since the static density variation is small (under 2\%), we will end up
neglecting it and proceed to linearize the equations, deriving
eq. \eqref{eqn:linCompEulerRepeat}--\eqref{eqn:pbc}  in section
\ref{sec:le}. 

This time there are two spatial coordinates, a horizontal coordinate $x$,
and a vertical coordinate $z$.
The (flat) bottom is $z=0$.
The moving top surface is $z = h(x,t)$.
The horizontal and vertical velocity components are $u$ and $w$ respectively,
and the water density is denoted by $\rho$.
The Euler equations are 
\begin{equation}
\begin{aligned}
\rho_t + (\rho u)_x + (\rho w)_z&= 0 \\[3pt]
(\rho u)_t + (\rho u^2 + p)_x + (\rho u w)_z &= 0 \\[3pt]
(\rho w)_t + (\rho u w)_x + (\rho w^2 + p)_z &=  -\rho g.
\label{eqn:compEuler}
\end{aligned}
\end{equation}
There is a ``no flow" boundary condition at the bottom boundary,
\begin{equation}
    w(x,z=0,t) = 0 .
\label{eqn:bottom}  
\end{equation}
The kinematic condition at the top boundary \cite{Whitham}
states that a particle that 
moves with the surface velocity stays on the surface,
\begin{equation}
       h_t + u h_x = w(x,h(x,t),t) \; .
\label{eqn:kinbc}  
\end{equation}
The dynamic boundary condition at the top is continuity of pressure,
\begin{equation}
       p(x, h(x,t), t) = p_{\mbox{\scriptsize \em atm}} + p_e(x,t) \; .
\label{eqn:dynbc}  \end{equation}
The left side of \eqref{eqn:dynbc} is pressure in the water evaluated at the top boundary.
The right side is the atmosphere's ambient pressure, which is the sum of the static
background atmospheric
pressure $p_{\text atm}$ and the dynamic blast wave overpressure $p_e(x,t)$.

For static solutions ($p_e=0$, $u=w=0$), the nonlinear 
equations (\ref{eqn:compEuler}) 
reduce to the hydrostatic balance condition 
\begin{equation}
      \frac{dp_0}{dz} = -g \rho_0(z) \; .
\label{eq:hb} \end{equation}
Let the water density  $\rho_w$ be the density at the water surface.
If the density differences are small (as they turn out to be), we may use
a linear approximation to the equation of state,
\[
        p(\rho) = p(\rho_w) + c_a^2 ( \rho - \rho_w) \; ,
\]
where $c_a$ is the acoustic sound speed in water at density $\rho_w$,
     $   c_a^2 = \frac{dp}{d\rho}(\rho_w) \; . $
The behavior of $\rho_0(z)$ is found by substituting this into (\ref{eq:hb}):
    $   \frac{d p_0}{dz} = c_a^2 \frac{d\rho_0}{dz} = -g \rho_0 \; . $
Therefore, for any two heights $z_1$ and $z_2$, we have
\[
       \rho_0(z_2) = \rho_0(z_1) e^{-\frac{g}{c_a^2}(z_2 - z_1)} \; .
\]
If $z_2 - z_1 = 4\,\mbox{km}$, and $c_a = 1500\, \frac{\mbox{m}}{\mbox{sec}}$, then 
     $  \frac{g}{c_a^2}(z_2 - z_1) < .02 \; . $
Therefore, the density varies by less than  about $2\%$
between the water surface and bottom.

We denote small disturbance quantities with a tilde, except for
the wave height response $h_r$, which we use for continuity with the
previous sections.
For example, the water density is $\rho_0(z) + \widetilde{\rho}(x,z,t)$.
These disturbances are driven by the atmospheric overpressure $p_e(x,t)$.
We substitute the expressions $\rho = \rho_0 + \widetilde{\rho}$, 
$u = \widetilde{u}$, $w = \widetilde{w}$ (since the velocities are linearized
around zero),  and $p = p_0 + c_a^2 \widetilde{\rho}$ into the 
Euler equations (\ref{eqn:compEuler}) and calculate up to linear terms in the disturbance variables.
Using the hydrostatic balance condition (\ref{eq:hb}), this gives
\[
\begin{split}
\widetilde{\rho}_t + \rho_0 \widetilde{u}_x + \rho_0 \widetilde{w}_z &= 0 \\[3pt]
\rho_0 \widetilde{u}_t + c_a^2 \widetilde{\rho}_x &= 0 \\[3pt]
\rho_0 \widetilde{w}_t + c_a^2 \widetilde{\rho}_z &= -\widetilde{\rho} g \; .
\end{split}
\]
Finally, we replace the (slightly) variable $\rho_0(z)$ with the constant $\rho_w$.
The resulting equations, which we use for analysis are
\begin{equation}
\begin{split}
\widetilde{\rho}_t + \rho_w \widetilde{u}_x + \rho_w \widetilde{w}_z &= 0 \\[3pt]
\rho_w \widetilde{u}_t + c_a^2 \widetilde{\rho}_x &= 0 \\[3pt]
\rho_w \widetilde{w}_t + c_a^2 \widetilde{\rho}_z &= -\widetilde{\rho} g \; .
\end{split}
\label{linCompEuler}
\end{equation}

The bottom boundary condition (\ref{eqn:bottom}) is already linear.
For the top boundary conditions, we express the water height as the sum of the 
background height $h_0$ and the disturbance height $h_r$:
\[
     h(x,t) = h_0 + h_r(x,t) \; .
\]
To leading order in $h_r$, $\widetilde{u}$ and $\widetilde{w}$,
the linear approximation to the kinematic boundary condition (\ref{eqn:kinbc}) is
\begin{equation}
    \frac{\partial{h_r}}{\partial t}(x,t) = \widetilde{w}(x,h_0,t) \; .
\label{eqn:linkin}  \end{equation}
For the dynamic boundary condition (\ref{eqn:dynbc}), which was
$ p(x,h(x,t),t) = p_{atm}+ p_e $,
we use the Taylor expansion and the perturbation approximation
\begin{align*}
       p(x,h(x,t),t) & \approx p(h_0) \,  + \, p_{0,z}(h_0) \,{h_r}(x,t)\\
                     & \approx p_0(h_0) + \widetilde{p}(x,h_0,t) +  p_{0,z}(h_0) \,{h_r}(x,t) \\
                 & \approx p_0(h_0) + c_a^2\widetilde{\rho}(x,h_0,t) +  p_{0,z}(h_0) \,{h_r}(x,t)\; .
\end{align*}
For the undisturbed quantities, the pressure at the top is 
$p(h_0) = p_{\mbox{\scriptsize \em atm}}$.
The hydrostatic balance relation (\ref{eq:hb}) in the water (applied at the top)
is $p_{0,z}(h_0) = - g \rho_w$.
Making these substitutions gives
\begin{equation}
p_0(h_0) + c_a^2 \widetilde{\rho} + p_{0,z} h_r = p_{atm} + p_e  ,
\end{equation}
giving the result
\begin{equation}
     c_a^2\widetilde{\rho}(x,h_0,t)- \rho_w  g \, {h_r}(x,t) = p_e (x,t) \; .
\label{eqn:lindyn}  \end{equation}
        
Summarizing, the linearized Euler equations are
\eqref{linCompEuler},  with  linearized boundary conditions \eqref{eqn:bottom},
\eqref{eqn:linkin} and \eqref{eqn:lindyn}.

\section*{Acknowledgments} 
We are particularly grateful to Mike Aftosmis, 
Oliver B\"uhler, and Randy LeVeque 
for more in-depth discussions.
It is a pleasure to thank our colleagues at the Courant Institute
for several lively discussions.  
We thank Michael Aftosmis
for providing the blast wave ground footprint model.
This effort was partially supported 
through a subcontract with Science and Technology
Corporation (STC) under  NASA Contract NNA16BD60C.

\bibliography{./references}

\begin{thebibliography}{10}
\providecommand{\url}[1]{{#1}}
\providecommand{\urlprefix}{URL }
\expandafter\ifx\csname urlstyle\endcsname\relax
  \providecommand{\doi}[1]{DOI~\discretionary{}{}{}#1}\else
  \providecommand{\doi}{DOI~\discretionary{}{}{}\begingroup
  \urlstyle{rm}\Url}\fi

\bibitem{aftosmis:bolide2016}
Aftosmis, M., Mathias, D., Nemec, M., Berger, M.: Numerical simulation of
  bolide entry with ground footprint prediction.
\newblock AIAA-2016-0998  (2016)

\bibitem{geoclaw:URL}
{G}eo{Claw} {W}eb {S}ite.
\newblock {\tt http://www.clawpack.org/geoclaw}.
\newblock \urlprefix\url{http://www.geoclaw.org/}

\bibitem{George:2008}
George, D.: Augmented {R}iemann solvers for the shallow water equations over
  variable topography with steady states and inundation.
\newblock J. Comp. Phys. \textbf{227}(6), 3089--3113 (2008)

\bibitem{noaa1}
Gica, E., Arcas, D., Titov, V.: Tsunami inundation modeling of {O}cean {S}hores
  and {L}ong {B}each, {W}ashington due to a {C}ascadia subduction zone
  earthquake.
\newblock Tech. rep., NOAA Center for Tsunami Research, Pacific Marine
  Environmental Laboratory (2014)

\bibitem{Gisler:2008}
Gisler, G.: Tsunami simulations.
\newblock Annu. Rev. Fluid Mech. \textbf{40}, 71--90 (2008)

\bibitem{GislerWeaverGitting:2010}
Gisler, G., Weaver, R., Gittings, M.: Calculations of asteroid impacts into
  deep and shallow water.
\newblock Pure Appl. Geophys. \textbf{168}, 1187--1198 (2010)

\bibitem{NTHMP}
Gonz{\'a}lez, F.I., LeVeque, R.J., Chamberlain, P., Hirai, B., Varkovitzky, J.,
  George, D.L.: Geoclaw model.
\newblock In: Proceedings and Results of the 2011 {NTHMP} Model Benchmarking
  Workshop, pp. 135--211. National Tsunami Hazard Mitigation Program, NOAA
  (2012).
\newblock
  \urlprefix\url{http://nthmp.tsunami.gov/documents/nthmpWorkshopProcMerged.pdf}

\bibitem{krakatoa}
Harkrider, D., Press, F.: The {K}rakatoa air-sea waves: an example of pulse
  propagation in coupled systems.
\newblock Geophys. J. R. Astr. Soc. \textbf{13}, 149--159 (1967)

\bibitem{krantzer:keller}
Kranzer, H., Keller, J.: Water waves produced by explosions.
\newblock J. Applied Physics \textbf{30}(3), 398--407 (1959)

\bibitem{article:LeVeque97}
LeVeque, R.: Wave propagation algorithms for multidimensional hyperbolic
  systems.
\newblock J.\ Comput.\ Phys. \textbf{131}, 327--353 (1997)

\bibitem{LeVequeGeorgeBerger:actaNumerica}
LeVeque, R., George, D., Berger, M.: Tsunami modelling with adaptively refined
  finite volume methods.
\newblock Acta Numerica pp. 211--289 (2011)

\bibitem{Liu:chap9}
Liu, P.: Tsunami modeling: Propagation.
\newblock In: E.~Bernard, A.~Robinson (eds.) The Sea: Tsunamis, vol.~15, pp.
  295--320. Harvard University Press (2009)

\bibitem{mandli:thesis}
Mandli, K.: Finite volume methods for the multilayer shallow water equations
  with applications to storm surges.
\newblock Ph.D. thesis, University of Washington (2011)

\bibitem{Melosh:2003}
Melosh, H.: Impact-generated tsunamis: An over-rated hazard.
\newblock In: 34th Lunar and Planetary Sciences Conference Abstract (2003)

\bibitem{MonserratVilibicRabinovich:2006}
Monserrat, S., Vilibic, I., Rabinovich, A.: Meteotsunamis: atmospherically
  induced destructive ocean waves in the tsunami frequency band.
\newblock Natural Hazards and Earth System Sciences \textbf{6}, 1035--1051
  (2006)

\bibitem{AGT:NASATM}
Morrison, D., Venkatapathy, E.: Asteroid generated tsunami: Summary of
  {NASA/NOAA} workshop.
\newblock Tech. Rep. {NASA/TM}-2194363, NASA Ames Research Center (2017)

\bibitem{petersenCramerFrankel:2002}
Petersen, M., Cramer, C., Frankel, A.: Simulations of seismic hazard for the
  {P}acific northwest of the {U}nited {S}tates from earthquakes associated with
  the {C}ascadia subduction zone.
\newblock Pure Appl. Geophys. \textbf{159}, 2147--2168 (2002)

\bibitem{Popinet:2015}
Popinet, S.: A quadtree-adaptive multigrid solver for the
  {S}erre-{G}reen-{N}aghdi equations.
\newblock J. Comp. Phys. \textbf{302}, 336--358 (2015)

\bibitem{Popova2013}
Popova, O., Jenniskens, P., Emelyanenko, V., Kartashova, A., Biryukov, E., {\em
  et al.}: Chelyabinsk airburst, damage assessment, meteorite recovery, and
  characterization.
\newblock Science \textbf{342}(6162), 1069--1073 (2013).
\newblock \doi{10.1126/science.1242642}

\bibitem{Proudman:1929}
Proudman, J.: The effects on the sea of changes in atmospheric pressure.
\newblock Geophysical Supplement to the Monthly Notices of the Royal
  Astronomical Society 2 \textbf{4}, 197--209 (1929)

\bibitem{UsluEtAl:2010}
Uslu, B., Eble, M., Titov, V., Bernard, E.: Distance tsunami threats to the
  ports of {L}os {A}ngeles and {L}ong {B}each, {C}alifornia.
\newblock Tech. rep., NOAA Center for Tsunami Research, Pacific Marine
  Environmental Laboratory (2010).
\newblock NOAA OAR Special Report, Tsunami Hazard Assessment Special Series (2)

\bibitem{vreugdenhil}
Vreugdenhil, C.: Numerical Methods for Shallow-Water Flow.
\newblock Kluwer Academic Publishers (1994)

\bibitem{WeissWunnemannBahlburg:2006}
Weiss, R., W\"unnemann, K., Bahlburg, H.: Numerical modelling of generation,
  propagation and run-up of tsunamis caused by oceanic impacts: model strategy
  and technical solutions.
\newblock Geophys. J. Intl \textbf{167}, 77--88 (2006)

\bibitem{Whitham}
Whitham, G.B.: Linear and nonlinear waves.
\newblock John Wiley \& Sons (1974)

\end{thebibliography}
\bibliographystyle{spmpsci}

\end{document}